\documentclass[a4paper,11pt]{article}
\pdfoutput=1

\usepackage{soul}
\usepackage{jcappub} 
\usepackage{tikz}
\usepackage{xcolor}

\usepackage[utf8]{inputenc}
\usepackage{amsmath, amsthm, mathtools}
\usepackage{braket}

\usepackage{tensor}
\usepackage{verbatim}
\usepackage{amsmath,amsfonts,amssymb}

\usepackage{fontawesome}

\usepackage{graphicx}
\usepackage{adjustbox}
\usepackage{tabularx}
\usepackage{tensor}
\usepackage{amsbsy}
\usepackage{braket}
\usepackage{slashed}
\usepackage{enumitem}
\hypersetup{%
,urlcolor=blue
,citecolor=blue
,linkcolor=blue
}

\usepackage[T1]{fontenc} 

\usepackage{braket}
\title{\boldmath Axion-Photon Conversion in 3D Media and Astrophysical Plasmas}

\author[a,c]{J. I. McDonald,}
\author[b]{B. Garbrecht,}
\author[c]{P. Millington}

\emailAdd{jamie.mcdonald@manchester.ac.uk}

\affiliation[a]{Centre for Cosmology, Particle Physics and Phenomenology\\
Université catholique de Louvain\\
Chemin du cyclotron, 2\\
Louvain-la-Neuve B-1348, Belgium}

\affiliation[b]{Technische Universität München\\
Physik-Department\\
James-Franck-Straße\\ 85748 Garching, Germany}

\affiliation[c]{
Department of Physics and Astronomy\\
University of Manchester\\
Manchester, M13 9PL, UK}

\abstract{With axions now a primary candidate for dark matter, understanding their indirect astrophysical signatures is of paramount importance. Key to this is the production of photons from axions in magnetised astrophysical plasmas. While simple formulae for axion-photon mixing in 1D have been sketched several decades ago, there has recently been renewed interest in robust calculations for this process in arbitrary 3D plasmas. These calculations are vital for understanding, amongst other things, the radio production from axion dark matter conversion in neutron stars, which may lead to indirect axion dark matter detection with current telescopes or future searches, e.g., by the SKA.  In this paper, we derive the relevant transport equations in magnetised plasmas. These equations describe both the production and propagation of photons in an arbitrary 3D medium due to the resonant conversion of axions into photons.  They also fully incorporate the refraction of photons, and we find no evidence for a conjectured phenomenon of dephasing. Our result is free of divergences that plagued previous calculations, and our kinetic theory description provides a direct link between ray tracing and the production mechanism. These results mark an important step toward solving one of the major open questions concerning indirect searches of axions in recent years, namely how to compute the photon production rate from axions in arbitrary 3D plasmas.
\bigskip

{\noindent \footnotesize This is an author-prepared post-print of \href{https://doi.org/10.1088/1475-7516/2023/12/031}{JCAP 12 (2023) 031}, published by IOP Publishing under the terms of the \href{https://creativecommons.org/licenses/by/4.0/}{CC BY 4.0} license. }

}

\begin{document}
\subheader{\hfill \parbox{5cm}{\raggedleft IRMP-CP3-23-25\\ TUM-HEP-1462/23}}

\maketitle
\flushbottom

\section{Introduction}
\label{sec:intro}

Astrophysical environments offer one of the most exciting arenas in which to probe physics beyond the Standard Model, providing extreme conditions in which to search for new laws of nature \cite{Baryakhtar:2022hbu}. Amongst these, observations of compact objects, namely stars and black holes, are opening new windows onto the Universe. Neutron stars in particular have become a source of increasing focus for the dark matter community, since they could provide indirect evidence for the existence of \textit{dark matter axions}. The question of what constitutes dark matter remains one of the most challenging open questions in basic science. Axions \cite{ref:PQ,ref:K,ref:SVZ,ref:DFSZ,ref:Zhit,ref:misalign1,ref:misalign2,ref:misalign3,Arvanitaki:2009fg,Svrcek:2006yi} are now becoming one of the most popular explanations. Indeed, axion dark matter has been the target of past, present and future laboratory searches \cite{DePanfilis:1987dk, Wuensch:1989sa, Hagmann:1990tj, ADMX:2009iij,ADMX:2018gho, ADMX:2019uok, ADMX:2018ogs, ADMX:2021nhd, ADMX:2021mio, Crisosto:2019fcj, CAPP:Lee_2020cfj, CAPP:Jeong_2020cwz, CAPP:Lee_2022mnc, CAPP:Kim2022, HAYSTAC:Brubaker2017, HAYSTAC:2018rwy, HAYSTAC:2020kwv, Alesini:2019ajt, Alesini:2020vny, McAllister:2017lkb, Quiskamp:2022pks, CAST:2020rlf, TASEH:2022vvu,
Grenet:2021vbb, Millar:2022peq, BREAD:2021tpx,Beurthey:2020yuq,
DMRadio:2022pkf, Marsh:2022fmo, McDonald:2021hus, Schutte-Engel:2021bqm}. While we await the development of the next generation of axion dark matter detectors, neutron stars continue to offer a powerful indirect probe in the axion mass range $m_\phi \simeq \mu eV$.

Dark matter axions can convert into radio photons in the strong magnetic fields of neutron stars \cite{pshirkov2009, hook2018,ref:NS-Japan, Hardy:2022ufh}. The conversion occurs in those regions where the axion and photon dispersion relations become degenerate. This occurs when $\omega_{\rm p} \simeq m_\phi$, where $\omega_p$ is the plasma mass. This mechanism has been the subject of extensive study \cite{Battye_2020,darling2020apj,Battye2022,foster2020,FosterSETI2022,Battye:2023oac}, spanning a range of radio observations. Efforts have also been made on the theoretical side to model the signal \cite{Leroy:2019ghm,Witte:2021arp,Battye:2021xvt,Millar2021,Carenza:2023nck}. Even so, one of the outstanding theoretical questions has been how to calculate the production process itself. The particular challenge is to understand the conversion from axions to photons in the limit in which the wavelengths of the two fields are much smaller than characteristic variational scales of the background plasma, so that they can be treated locally as particle states in the spirit of the WKB approximation. Whilst the classic reference \cite{Raffelt:1987im} has become the go-to work for describing axion-photon mixing in this limit, it deals only with mixing between axions and photons propagating in a single direction, making it suitable only for describing 1D problems. Reference~\cite{hook2018} employed a heuristic description of resonant axion-photon mixing in plasmas, but this too was essentially a 1D description. It was soon noted \cite{Battye_2020} that 3D effects would be important in describing the production.

There are a number of challenges when undertaking the calculation in 3D. First, the large hierarchy of scales in the WKB limit makes solving the problem numerically difficult. Second, even if the problem could be solved numerically, in order to implement ray-tracing strategies \cite{Witte:2021arp,Battye2022}, the conversion probability must be computed millions of times for each parameter choice of neutron star environment and axion mass. It is therefore crucial to derive an analytic expression for the production of photons from axions in arbitrary 3D media.

To derive an analytic expression for the photon production rate, one must address a number of issues. First, due to the vector nature of photons, multiple polarisations of the electric field couple to one another in Maxwell's equations in anisotropic media.  Furthermore, one has to recover a particle and phase-space picture, starting from a set of classical wave equations, namely those of axion electrodynamics.  This requires one to somehow identify worldlines of photons and integrate along these to obtain the intensity of photons due to axion conversion. Next, one must have a systematic calculation scheme for exploiting the hierarchy of scales between axion and photon wavelengths, and much larger background scales. This entails performing an expansion in gradients along the lines of a WKB approximation. An additional challenge is that axions and photons in general have different dispersion relations away from the resonance, which complicates finding classical solutions in terms of plane waves. This issue was raised  already in Ref.~\cite{Battye_2020}.

  One also has to deal with the fact that plasma modes are not vacuum photons, but rather collective plasma excitations of charge carriers and the electromagnetic field. This must be incorporated into any photon production rate. In addition, owing to the 3D nature of astrophysical plasmas, one has to incorporate the fact that photons will be refracted relative to axions, and therefore the two particles move on different paths as they enter and exit the resonant point. This fact was noted in Ref.~\cite{Witte:2021arp}, which queried the extent to which photon refraction may affect the conversion probability. Finally, one should ensure that any divergences appearing in the conversion probability are regulated by appropriate physical quantities appearing in the phase-space integration over axion and photon momenta, without the need to impose arbitrary regulatory cutoffs.

Reference~\cite{Millar2021} approached the calculation of axion-photon conversion in 3D magnetised plasmas by directly solving the equations of axion electrodynamics,
seeking plane-wave solutions in the WKB limit to derive a first-order transport-like equation for the electric field amplitude sourced by axions. This equation was then solved using a stationary phase approximation at the resonance to derive the outgoing electric field of the photon, which was subsequently used to define a rate for axion-photon conversion.

In this work, rather than attempting to solve classical field equations directly, we show instead that kinetic theory provides a simple, elegant and powerful framework, which offers a consistent language to address all the issues mentioned above. In this formalism, the calculation can be carried out in a controlled way by deriving a Boltzmann-like transport equation for the phase-space distribution of photons and axions, denoted $f_\gamma$ and $f_\phi$, respectively. The pertinent Boltzmann-like equation is 
\begin{equation}\label{eq:BoltzmannExplicitIntro}
   \partial_k \mathcal{H} \partial_x f_\gamma	- \partial_x \mathcal{H} \partial_k f_\gamma  =   g_{a \gamma \gamma}^2 E_\gamma^2 \big| \textbf{B} \cdot \boldsymbol{\varepsilon} \big|^2 2 \pi \delta \left(E_\gamma ( \textbf{k},x)^2 - E_\phi(\textbf{k})^2 \right) f_\phi \, ,
\end{equation}
where $\mathcal{H}$ is the Hamiltonian describing the propagation of photons in the medium in question, and $E_\phi$ and $E_\gamma$ are the axion and photon energies, respectively. Here, the delta function in the collision term on the right-hand side imposes the kinematic condition for the conversion to take place. The remaining terms in the collisional piece are the magnetic field $\textbf{B}_{\rm ext}$, the unit photon polarisation 3-vector $\boldsymbol{\varepsilon}$ and the axion-photon coupling $g_{a \gamma \gamma}$.

Equation~\eqref{eq:BoltzmannExplicitIntro} describes the production of photons due to axions and their subsequent propagation along photon worldlines. By solving the associated characteristics of the Liouville-Vlasov operator that appears on the left-hand side, which are nothing more than Hamilton's equations
\begin{equation}\label{eq:HamiltonEqsIntro}
   \frac{{\rm d} x^\mu (\lambda)}{d \lambda}   = \frac{\partial  \mathcal{H}}{\partial k_\mu}\;, \qquad \frac{{\rm d} k_\mu (\lambda)}{d \lambda}   =-  \frac{\partial  \mathcal{H}}{\partial x^\mu} \, ,
\end{equation}
one can compute the solution for the phase-space distribution of photons $f_\gamma$ along these integral curves, thereby obtaining the ratio 
\begin{equation}\label{eq:PIntro}
    P_{a \gamma} = f_\gamma/f_\phi
\end{equation}
at the point of conversion. In general, this expression depends on the kinematics of the photon and axion at that point of conversion. One medium of particular relevance is the strongly magnetised plasma, where the cyclotron frequency greatly exceeds the photon and plasma frequencies. For this case, we obtain an explicit expression 
\begin{equation}\label{eq:PStrongBIntro}
	P_{a \gamma } = \frac{\pi}{2} \frac{ g_{a \gamma \gamma}^2\left| \textbf{B}_{\rm ext} \right|^2E_\gamma ^4 \sin ^2 \theta }{\cos ^2 \theta
   \, \omega _p^2 \left(\omega _p^2-2 E_\gamma ^2\right)+E_\gamma ^4} 	\frac{1}{\left| \textbf{v}_p \cdot \nabla_\textbf{x} E_\gamma  \right|}\;.
\end{equation}
The formula \eqref{eq:PStrongBIntro} is the central result of our paper, describing the production of photons from axion dark matter in the plasma surrounding neutron stars. We stress, however, that the description above applies equally to mixing between arbitrary particles in any medium.  A large part of this paper is therefore dedicated to justifying, on the basis of a first-principles calculation that the form of Eq.~\eqref{eq:BoltzmannExplicitIntro} is correct. Our scheme also allows one to make direct contact with ray-tracing routines \cite{Battye:2021xvt,Witte:2021arp,McDonaldInPrep,Tjemsland:2023vvc}. Not only does this make clear the physical interpretation at every step of the calculation, but the final result is well-behaved when integrated over the photon emission surface, and free from divergences in a formal sense, so long as perturbation theory remains valid.

To derive a consistent set of transport equations from which Eq.~\eqref{eq:BoltzmannExplicitIntro} emerges, we use a first-principles approach formulated within the Schwinger-Keldysh closed time path formalism~\cite{Schwinger:1960qe, Keldysh:1964ud}.  This framework provides a path-integral representation for the generating functional of statistical expectation values and, in this way, provides a means for us to extract the correlation functions in the plasma.  The Schwinger-Dyson equations for these Green's functions can then be derived from the corresponding effective action, in particular, the so-called two-particle irreducible effective action~\cite{Cornwall:1974vz} and its generalisation in the closed time path formalism~\cite{Calzetta:1986cq}.  From there, and following approaches similar to those introduced by Kadanoff and Baym~\cite{Baym:1961zz}, and Danielewicz~\cite{Danielewicz:1982kk}, the Schwinger-Dyson equation can be used to derive evolution equations for phase-space distribution functions, and it is these that will allow us to describe the axion-photon mixing in the magnetised plasma.  The above approaches to non-equilibrium quantum field theory and transport phenomena have found wide application, particularly in the context of leptogenesis (for reviews, see Refs.~\cite{Dev:2017trv, Dev:2017wwc} and references therein) and baryogenesis~\cite{Prokopec:2003pj, Prokopec:2004ic} (both describing the generation of the baryon asymmetry of the universe), and in analyses of neutrino quantum kinetics~\cite{Vlasenko:2013fja, Blaschke:2016xxt} (e.g., in core collapse supernovae).

By projecting the resulting transport equations onto photon worldlines, we are able to derive a simple expression for the conversion probability of axions and photons in terms of a differential equation in a single integration parameter along photon trajectories. In this way, we obtain a description of photon production from axions in a magnetised plasma, taking us closer to a solution of one of the major open theoretical questions in indirect radio searches for axions.

The structure of this paper is as follows. In section~\ref{sec:SchwingerDyson}, we use first-principles techniques from non-equilibrium quantum field theory to derive transport equations for photons that describe their production and propagation in arbitrary media, and then use these equations to derive the explicit form of the photon source term arising from axions. In section \ref{sec:PhotonProuction}, we then employ a method of characteristics to solve the transport equation along the worldines of photons, which allows us to derive a conversion probability for the production of photons from axions. Finally, we apply these results to a strongly magnetised plasma, thereby obtaining the central result of our paper. This is, in fact, extendable to mixing with photons and other particles in general 3D media, as demonstrated in section \ref{sec:parityEven} for the example of a parity-even scalar. Finally, in section \ref{sec:conclusions}, we offer our conclusions. Further technical details are provided in the appendices.

\section{Kinetic Equations}\label{sec:SchwingerDyson}

To arrive at kinetic equations that describe the evolution of the photon phase-space densities $f^c$ (with $c$ labelling different polarisations), we are primarily interested in the Schwinger-Dyson equations that determine the photon Wightman propagators $D^{>}_{\mu\nu}(x,y)=\braket{A_{\mu}(x)A_{\nu}(y)}_{\rho}$ and $D_{\mu\nu}^<(x,y)=\braket{A_{\nu}(y)A_{\mu}(x)}_{\rho}$, as expressed in the Heisenberg picture. The subscript $\rho$ on the expectation values indicates crucially that these are statistical expectation values of the vector fields, weighted by the density operator $\rho$, which, in the Heisenberg picture, is time-independent and encodes the initial state of the ensemble. The relevant Schwinger-Dyson equations can be derived from first principles within the so-called Schwinger-Kelydsh closed time path formalism by means of the two-particle irreducible effective action. The technical details of these frameworks can be found in Refs.~\cite{Berges:2004yj, Blaizot:2001nr}. However, it is sufficient for our purposes to begin with the Schwinger-Dyson equation for the Wightman function \cite{Lin2019,Blaizot:2001nr}
\begin{align}\label{eq:Wightmann1}
\mathcal{D}_0^{\mu \nu}  D^{<,>}_{\nu \sigma}(x, y)=  \int {\rm d}^4z\; [\Pi_{\rm R}^{\mu \nu}(x, z) D^{<,>}_{\nu \sigma}(z, y) + \Pi^{<,> \mu \nu}(x, z) D_{A\, \nu \sigma}(z, y)],
\end{align}
where $D_{A\, \nu \sigma}(z, y)$ is the advanced photon propagator and
\begin{equation}\label{eq:DmunuSpatial}
	\mathcal{D}_0^{\mu \nu} = \partial^2 g^{\mu\nu}-\partial^\mu\partial^\nu+\frac{1}{\xi}P^{\mu\alpha}P^{\nu\beta}\partial_\alpha\partial_\beta,
\end{equation}
whose partial derivatives are understood to act on the first spacetime coordinate $x$ of the Wightman propagators $D^{<,>}_{\nu \sigma}(x, y)$. The projection operator $P^{\mu \alpha}$ is defined by $P^{\mu \alpha} = u^\mu u^\alpha - \eta^{\mu \alpha}$, where $u^\mu$ is the 4-velocity of the plasma and $\eta^{\mu \alpha}$ is the Minkowski metric. The term proportional to $1/\xi$ is a covariant gauge-dependent term. $\Pi^{\mu\nu}_{\rm R}(x,z)$ is the retarded photon self-energy, and $\Pi^{<,>\mu\nu}(x,z)$ are the Wightman self-energies (which are related to the absorptive parts of the usual Feynman self-energy).  These self-energies contain contributions from both axions and the background plasma.  In the present setup, the contribution arising from the plasma will give the dominant contribution to the photon dispersion relation, while the piece arising from axions will give a production term for photons.

For a given two-point function or self-energy $B = D,\, \Pi$, it is helpful to define a ``Hermitian'' function
\begin{align}\label{eq:AppHermitianDef}
    B_{\rm H}=\frac12\left(B_{\rm R}+B_{\rm A}\right),
\end{align}
where the subscripts $\rm R$ and $\rm A$ refer to the corresponding retarded and advanced functions, respectively. Noting that
\begin{equation}
B_{\rm R}-B_{\rm A}=i\left(B^>-B^<\right),
\end{equation}
we can then write
\begin{equation}
    B_{\rm R(A)}=B_{\rm H}+(-)\frac{i}{2}\left(B^>-B^<\right)
\end{equation}
and recast Eq.~\eqref{eq:Wightmann1} in the form of a Kadanoff-Baym equation
\begin{align}\label{eq:KB1}
    &\mathcal{D}_0 \cdot D^{<,>}(x,y) - \int {\rm d}^4 z\; \Pi_{\rm H}(x,z)\cdot D^<(z,y)-\int {\rm d}^4 z \, \Pi^{<,>}(x,z) \cdot D_{\rm H}(z,y)\notag\\
    &\qquad =
    -\frac{i}{2}\int  {\rm d}^4 z  \left[\Pi^>(x,z)\cdot D^<(z,y)-\Pi^<(x,z)\cdot D^>(z,y)\right],
\end{align}
wherein we can already identify the collision terms on the right-hand side.

Ultimately, we would like to derive a governing equation for photon distribution functions $f_\gamma$, which are densities in phase space. It therefore proves convenient to move from configuration space $(x,y)$ to Wigner space $(k,X)$, wherein we Fourier transform with respect to a relative coordinate $s^{\mu}=x^{\mu}-y^{\mu}$ to introduce the four-momentum $k$ and leave behind a dependence on the central coordinate $X^{\mu}=(x^{\mu}+y^{\mu})/2$. In doing so, the convolution integral over the spacetime coordinate $z$ in Eq.~\eqref{eq:KB1} is traded for an infinite series of derivatives in the Wigner-space coordinates $(k,X)$. Further details are provided in Appendix~\ref{App:kinetic}.

When the wavelength of quasiparticle excitations is short compared to background gradients, such that $\partial_X \cdot \partial_k \ll 1$, we can truncate this series of gradients to first order\footnote{There are situations where certain gradients should be summed up to all orders to capture the finite width of excitations or to resolve mass shells within flavour oscillations~\cite{Garbrecht:2011xw,Fidler:2011yq}.} to obtain
\begin{align}
\label{eq:firstordergrad}
   &\overline{\cal D}_0\cdot D^{<,>}-\Pi^H\cdot D^{<,>} -\frac{i}{2} \{\Pi^H, D^{<,>}\}
    -\Pi^{<,>}\cdot D^H  -\frac{i}{2}\{\Pi^{<,>}, D^H\}\nonumber\\&\qquad=-\frac{i}{2}\left(\Pi^{>}\cdot D^{<}-\Pi^{<}\cdot D^{>}\right)-\frac{1}{4}\left(\{\Pi^<,D^>\}-\{\Pi^>,D^<\}\right),
\end{align}
wherein
\begin{align}
\overline{\mathcal{D}}_0^{\mu\nu}(k,X)={\cal D}_0^{\mu\nu}(k)+
    \frac i2 \frac{\partial}{\partial k^\alpha}  {\cal D}_0^{\mu\nu}(k)\partial^\alpha_X -\frac18 \frac{\partial}{\partial k^\alpha} \frac{\partial}{\partial k^\beta}{\cal D}_0^{\mu\nu}(k)\partial^\alpha_X \partial^\beta_X,
\end{align}
with
\begin{align}
    \mathcal{D}_0^{\mu\nu}(k)=-k^2 \eta^{\mu\nu}+k^\mu k^\nu -\frac1\xi P^{\mu\alpha}P^{\nu\beta}k_\alpha k_\beta.
\end{align}
The terms at first-order in gradients involve Poisson brackets of the form
\begin{equation}\label{eq:PB}
\left\{
A, B \right\}
= \partial_k A \cdot \partial_X B - \partial_X A \cdot \partial_k B.
\end{equation}
The kinetic equation is obtained from the anti-Hermitian part of Eq.~\eqref{eq:firstordergrad}, leading to
\begin{align}
    \label{eq:kineticmain}
    \bigg[{\cal D}
    ,D^{<,>}\bigg]_-
    +\frac i2 \bigg[\partial_{[k} {\cal D}, \partial_{X]} D^{<,>}\bigg]_+
    =&-\frac i2 \bigg[\Pi^>,D^<\bigg]_++\frac i2 \bigg[\Pi^<,D^>\bigg]_+,
\end{align}
where $\left[A,B\right]_{\mp}=A\cdot B\mp B\cdot A$, square brackets on the Wigner coordinates $k$ and $X$ indicate antisymmetrisation via $\partial_{[k}A \partial_{X]} B = \partial_{k} A \partial_X B- \partial_{X} A \partial_{k} B$,
and we have defined
\begin{equation}\label{eq:DRInverse}
    \mathcal{D}^{\mu \nu}(k,X) =  -k^2g^{\mu\nu} + k^\mu k^\nu - \frac{1}{\xi} P^{\mu\alpha} P^{\nu\beta} k_\alpha k_\beta - \Pi_{_{\rm H}}^{\mu \nu}.
\end{equation}
The self-energy $\Pi_{\rm H}$ is the Hermitian part of the retarded self-energy, which can be obtained according to Eq.~\eqref{eq:AppHermitianDef}, and it can be read off from the (loss-free) classical photon polarisation tensor. We note that we have neglected gradient corrections to the collision terms on the right-hand side of the kinetic equation~\eqref{eq:kineticmain}, which do not contribute to the outgoing photon flux at second order in the axion-photon coupling, as described further in Appendix~\ref{App:kinetic}.

To leading order in gradients, Eq.~\eqref{eq:DRInverse} defines the inverse of the retarded Wigner function $D_{\rm R}$, such that 
\begin{equation}\label{eq:RetardedEOM}
 \mathcal{D}^{\mu\nu}(k,X) D_{\nu\rho}^R(k,X)   = -\delta^\mu_\rho.
\end{equation}
Of particular importance are the eigenvectors $\varepsilon^b$ and eigenvalues $\mathcal{H}^b$ of $\mathcal{D}$, where Latin characters $a,b,c,\dots$ index the eigenbasis. We choose the label $\phi$ for axions to avoid confusion with the label $a$. The eigenvectors satisfy
\begin{equation}\label{eq:OrthonormalBasis}
	\mathcal{D}\cdot \varepsilon^c  = \mathcal{H}^c \, \varepsilon^{c }, \qquad  \varepsilon^{b \,*} \cdot \varepsilon^{c} = \delta^{ b c}, \qquad \sum_c \varepsilon^c_\mu \varepsilon^{ c \, *}_\nu = -\eta_{\mu \nu} \, .
\end{equation}

The $\varepsilon^c$ give the physical polarisations of the eigenmodes, and setting $\mathcal{H}_c(k,X) =0$ gives the dispersion relation of the mode. The notation $\mathcal{H}_c$ for the eigenvalues is suggestive, since, as we will see in subsequent sections, they will become the Hamiltonians that describe the propagation of each eigenmode.

By projecting Eq.~\eqref{eq:kineticmain} onto the polarisation basis vectors, we can derive a kinetic equation that describes the transport of the plasma mode associated to $\varepsilon^a$. The full details are given in Appendix \ref{App:kinetic}. Explicitly, we find
\begin{equation}\label{eq:KineticEq}
	\varepsilon^{a*}\cdot \left\{ \mathcal{D}, D^<  \right\}  \cdot\varepsilon^a  
 + \text{H.c.}
 = \varepsilon^{a*}\cdot\left(D^> \Pi^< - \Pi^> D^< \right)\cdot\varepsilon^a + \text{H.c.}\, ,
\end{equation}
where H.c.\ indicates the addition of the Hermitian-conjugate term. Equation \eqref{eq:KineticEq} generalises the results of Sec.~2.3.2 of Ref.~\cite{Blaizot:2001nr} for scalars to Abelian gauge fields (see also Ref.~\cite{Lin:2021mvw}). 

As explained in greater detail in Appendix~\ref{App:kinetic}, we only consider \textit{dispersive} effects from the plasma, as embodied by the Hermitian part of the photon response function $\Pi_{\rm H}$ in Eqs.~\eqref{eq:DRInverse} and \eqref{eq:KineticEq}. This gives the mass-shell condition for photons and captures refractive effects represented by the Poisson bracket $\left\{ \mathcal{D},D^< \right\}$. We do not consider \textit{absorptive} plasma contributions to photons arising from $\Pi^{<,>}$, which correspond to scattering, production or absorption of photons due to real processes that involve scatterings with charge carriers in the plasma. Hence, we only consider axion contributions to the gain and loss terms $\Pi^{<,>}$ for photons appearing on the right-hand side of Eq.~\eqref{eq:KineticEq}.

Since we are interested here in the production of photons due to axions, this approximation is justified provided the mean free path of photons is long compared to the length scales over which axions resonantly convert to photons. This is equivalent to neglecting the attenuation coefficients in the standard equations for radiative transfer \cite{Zheleznyakov1969, Befki1966}. Of course, in the context of neutron stars, attenuation via the cyclotron resonance can be significant \cite{Witte:2021arp} for more strongly magnetised stars. However, this occurs in regions far from where axion-photon conversion takes place. We therefore do not consider absorptive plasma effects in this work, since they are not relevant for perturbative axion-photon production in the local environments that we have in mind.  

To proceed, we require a solution for $D^<$ that involves the photon distribution functions.  We begin by noting that, since $\mathcal{D}$ is self-adjoint, we can express $D_{\rm R}$ in terms of the eigenvectors and their inverse eigenvalues via Eqs.~\eqref{eq:RetardedEOM} and \eqref{eq:OrthonormalBasis} as
\begin{equation}\label{Eq:DR}
	D_{{\rm R} \, \mu \nu}(k,X) = \sum_{c} \frac{ \varepsilon^{c}_\mu \varepsilon^{c\, *}_\nu }{\mathcal{H}_c + i \eta \sigma(k_0)}.
\end{equation}
The pole prescription $i\eta\sigma(k_0)$ involves the sign function $\sigma(k_0)$, which ensures the pole structure corresponding to physical states is consistent with the causal properties of the retarded function.
Any solutions $D^{<,>}$ for the Wightman functions must satisfy
\begin{equation}\label{eq:SpectralFunction}
	\rho(k,X) = D^>(k,X) - D^<(k,X) =  - 2 \,{\rm{Im} }D_{\rm R}(k,X),
\end{equation}
where $\rho$ is the spectral function. We can then use Eqs.~\eqref{Eq:DR} and~\eqref{eq:SpectralFunction} to express the spectral function as
 \begin{equation}\label{eq:spectralDecomp}
 \rho_{\mu \nu} = \sum_{c} \varepsilon^{c}_\mu \varepsilon^{c\, *}_\nu   2 \pi \sigma (k_0)\delta(\mathcal{H}_c).
 \end{equation}

We are now in a position to seek Ansätze for the Wightman functions $D^\gtrless$, which must respect Eqs.~\eqref{eq:SpectralFunction} and \eqref{eq:spectralDecomp}. We consider solutions of the form
\begin{subequations}
\label{eq:SpectralAnsatz}
\begin{align}
D_{\mu \nu}^< &=  \sum_{c} 2 \pi\varepsilon^{c}_\mu \varepsilon^{c \, *}_\nu  \left\{ \theta(k_0)f^c(k,X) +\theta(-k_0)\left[1+f^c(-k,X)\right]\right\}\delta(\mathcal{H}_c), \label{eq:WightmannAnsatz1}\\ 
D_{\mu \nu}^> &=  \sum_{c} 2 \pi\varepsilon^{c}_\mu \varepsilon^{c \, *}_\nu\left\{\theta(k_0)\left[1+f^c(k,X)\right]+ \theta(-k_0)f^c(-k,X)\right\}\delta(\mathcal{H}_c).
\label{eq:WightmannAnsatz2}
\end{align}
\end{subequations}
Evidently, Eq.~\eqref{eq:SpectralAnsatz} satisfies Eq.~\eqref{eq:SpectralFunction}. The $f^c = f^c(k,X)$ are the distribution functions of quasiparticles (e.g., plasma modes) within the medium.
Note that, in order for the $f^c$ to be interpreted as distribution functions, each $\mathcal{H}_{c}\equiv \mathcal{H}_c(k_0,\textbf{k})$ must correspond to a single dispersion relation, so that, from $\mathcal{H}_c(k_0,\textbf{k})  = 0 $, one obtains an equation $k_0^2 = E(\textbf{k})^2$, where $E(\textbf{k})$ gives the energy of a physical eigenmode in the system. This can be obtained from the Hermitian part of Eq.~\eqref{eq:firstordergrad}, as shown in Appendix~\ref{App:kinetic}. For our purposes, we will need only the positive roots $k_0>0$ to which the analysis that follows will be restricted. Were any of the $\mathcal{H}_c$ to correspond to multiple physical states, this would spoil the interpretation of the $f^c$ as corresponding to distinct eigenmodes within the system. We do not offer a general proof that this is always true for the decomposition in Eq.~\eqref{Eq:DR} (though on physical grounds it seems such arguments must surely exist). Instead, we have verified, for instance, for a strongly magnetised plasma, that there are precisely three physical modes, and that each one corresponds to the zeros of a different $\mathcal{H}_c$. This is explained in Appendix \ref{App:Eigenmodes}, where we analyse the spectral structure of the operator \eqref{eq:DRInverse} in greater detail.  Similarly for an isotropic, cold, unmagnetised plasma, we find that there are modes $k_0 = 
\left| \textbf{k} \right|$ and $k_0 = 
\smash{\sqrt{|\textbf{k}|^2 + \omega_p^2}} $, and each of these corresponds to zeros of distinct $\mathcal{H}_c$.

We remark in passing that the summands in Eq.~\eqref{eq:SpectralAnsatz} are diagonal in polarisation labels. Formally, one could imagine a more general Ansatz that incorporates quantum correlations between different polarisation eigenstates, which would schematically involve a sum of the form $D^{\delta}_{\mu\nu} = \sum_{c c'} \varepsilon^c_{\mu} \varepsilon^{c'*}_{\nu} f^{c c'}(k,X)\delta(\mathcal{H}_{c c'}) $ (again taking only the $k_0>0$ part). Here, the $f^{c c'}$ with $c \neq c'$ would quantify the size of off-diagonal quantum correlations between different polarisations. Similarly, $\mathcal{H}_{c c'}$ gives an effective mass-shell condition of off-diagonal correlations. This typically gives a mass-shell condition that is an averaging of the mass-shell conditions for the diagonal states $c$ and $c'$, as was shown in Ref.~\cite{Prokopec:2003pj,Kartavtsev:2015vto,Battye_2020}. This form is analogous to the generalised density matrix describing quantum correlations between different neutrino flavours (see, e.g., Refs.~\cite{Sigl:1993ctk,Stirner:2018ojk}).

To generate off-diagonal correlations, symmetries in the theory must be broken in such a way that polarisation eigenstates (or flavour, in the case of neutrinos) are not conserved. For photons in a magnetised plasma, since different polarisations have different dispersion relations, mixing is in general kinematically forbidden. The exception would be if a  background field can inject momentum transfer between two different modes, or if there is a level-crossing in which the dispersion relations of different polarisations become degenerate, allowing for resonant transitions to occur. The latter is discussed briefly in Appendix \ref{App:Eigenmodes} after Eqs.~\eqref{eq:CharactersticMoments} and \eqref{eq:CharactersticMoments2}. 
Assuming that none of these special cases occur at the point of axion-photon conversion, off-diagonal correlations will not be generated and our diagonal Ansatz \eqref{eq:SpectralAnsatz} is sufficient to characterise the problem at hand. This should fully capture classical physics encoded in axion electrodynamics. We therefore leave a deeper treatment of off-diagonal quantum-correlations for future work.

Returning to the derivation of the transport equations, we treat the quasiparticles as being stable, with real mass-shell relations given by the zeros of $\mathcal{H}_c$. Inserting then the Ansatz
\eqref{eq:SpectralAnsatz} 
into Eq.~\eqref{eq:KineticEq}, we arrive at
\begin{equation}\label{eq:kinetic}
\varepsilon^{c \, * }_\rho \Big\{ [\mathcal{D}]\tensor{}{^\rho ^\mu} , \sum_{b} \varepsilon^b_\mu \varepsilon^{*b}_\nu
f_b \delta(\mathcal{H}_b) \Big\} \varepsilon^{c \, \nu}=
 \varepsilon^{c \, *} \cdot \Big( (1+ f_c) \Pi^<  - \Pi^> f_c\Big) \cdot \varepsilon^c
\delta(\mathcal{H}_c).
\end{equation}
 After some relatively straightforward manipulations of the basis vectors and their derivatives (see Appendix \ref{App:Poisson}), the left-hand side of this equation can be simplified to give
\begin{equation}\label{Eq:BoltzmannFull}
\delta(\mathcal{H}_c) \Big\{ \mathcal{H}_c,
f_c  \Big\} =
  \varepsilon^{c \, *} \cdot \Big( (1+ f_c) \Pi^<  - \Pi^> f_c\Big) \cdot \varepsilon^c \delta(\mathcal{H}_c) \, . 
\end{equation}
The left-hand term is now recognisable as the standard Vlasov operator \cite{Blaizot:2001nr} 
\begin{equation}\label{eq:Liouville}
    \Big\{ \mathcal{H}_c,
f_c  \Big\} = \partial_k \mathcal{H} \cdot \partial_X f_c	- \partial_X \mathcal{H} \cdot \partial_k f_c,
\end{equation}
as appears in the collisionless Boltzmann equation. Furthermore, the notation for $\mathcal{H}_c$ now becomes apparent:\ it is nothing more than the Hamiltonian for the modes. Finally, the terms on the right-hand side of Eq.~\eqref{Eq:BoltzmannFull} are the collision terms, with the first and second terms corresponding to the gain and loss processes for photons, respectively, as is apparent from the quantum statistics factors $f_c$ and $(1+f_c)$, with the latter corresponding to Bose enhancement of the production process. This is precisely a quantum Boltzmann or transport equation, which governs the evolution of the phase-space distribution $f_c$ of each quasiparticle photon mode.

The interpretation of the $\varepsilon^a\equiv \varepsilon^{a}(k,X)$ as polarisation vectors readily follows by considering the equation of motion for the one-point function
\begin{equation}
    \int {\rm d}^4 z\; \mathcal{D}^{\mu\nu}(x,z)A_\nu(z) = 0,
\end{equation}
or more explicitly, from Eqs.~\eqref{eq:Wightmann1}-\eqref{eq:DmunuSpatial}, 
\begin{align}\label{eq:Classical}
\left(\partial^2g^{\mu\nu}-\partial^\mu\partial^\nu+\frac{1}{\xi}P^{\mu\alpha}P^{\nu\beta}\partial_\alpha\partial_\beta \right) A_\nu(x)- \int {\rm d}^4z\; \Pi^{\mu\nu}_{\rm H}(x,z)A_\nu(z) = 0.
\end{align}
In the linear regime, wherein we can neglect any terms in $\Pi_{\rm H}^{\mu\nu}$ that depend on the one-point function, we see that $A_{\mu}$ is the zero mode of the same operator $\mathcal{D}$ that appears in the Schwinger-Dyson equation for the Green's functions. In the homogeneous limit $\Pi_{\rm H}(x,z) = \Pi^{\rm hom}_{\rm H}(x - z)$, the classical solution $A_{\mu}$ can therefore be decomposed in momentum space in terms of the homogeneous polarisation vectors $\varepsilon^a(k)$ that satisfy the Fourier transformed eigenvalue problem
\begin{align}\label{eq:PolarisationVectors}
\left[-k^2g^{\mu\nu}+k^\mu k^\nu-\frac{1}{\xi}P^{\mu\alpha}P^{\nu\beta}k_\alpha k_\beta - \Pi^{\rm hom \, \mu \nu }_{H}(k)\right] \varepsilon^a_\nu(k) = 0,
\end{align}

Thus, in the weak gradient expansion, i.e., the WKB limit, the response function $\Pi_{\rm H}(k,X)$ should locally resemble that of an infinite homogeneous medium $\Pi_{\rm H}^{\rm hom}(k)$ with constant properties, e.g., constant plasma frequency $\omega_p$, and magnetic field $\textbf{B}$. This means that we can identify  $\Pi_{\rm R}(k,X) = \left. \Pi^{\rm hom}_{\rm R}(k)\right|_{\textbf{B} \rightarrow \textbf{B}(X), \omega_p \rightarrow \omega_p(X)} $ and promote the homogeneous polarisation vectors $\varepsilon^a(k)$ to \textit{inhomogeneous} polarisation vectors $\varepsilon^a(k,X)$ that satisfy
\begin{align}
\left[-k^2g^{\mu\nu}+k^\mu k^\nu-\frac{1}{\xi}P^{\mu\alpha}P^{\nu\beta}k_\alpha k_\beta - \Pi^{\rm\mu \nu }_{H}(k,X)\right] \varepsilon^a_\nu(k,X) = 0,
\end{align}
in accordance with the Ansatz for the retarded propagator in Eq.~\eqref{Eq:DR}.

To make the physical properties of our transport equation more manifest, it is informative to integrate Eq.~\eqref{Eq:BoltzmannFull} over positive frequencies by acting with $\int_0^\infty {\rm d}k_0 $. This leads to an explicit form
\begin{equation}\label{eq:BoltzmannTime}
\partial_t f_\gamma + \textbf{v}_g \cdot \nabla_\textbf{X} f_\gamma - \nabla_\textbf{X} E_\gamma \cdot \nabla_\textbf{k} f_\gamma - \partial_t E_\gamma \partial_{E} f_\gamma = \frac{ \varepsilon^*_\mu \varepsilon_\nu}{  \partial_{k_0} \mathcal{H} } \Big( (1+ f_\gamma) \Pi^{<\, \mu \nu}  - \Pi^{>\, \mu \nu} f_\gamma\Big)  \, ,
\end{equation}
where we have dropped the subscript $c$ labelling the photon mode and have written the photon distribution function for a given mode  as $f_\gamma$ to distinguish it from the axion distribution function $f_\phi$ to be introduced later. Note that we use the shorthand notation $\partial_E f_\gamma \equiv \lim_{k_0 \rightarrow E_\gamma(\textbf{k},X)}\partial_{k_0} f_\gamma(k_0, \textbf{k}, X)$. We have also used $\delta(\mathcal{H}_c) =  \delta(k_0- E_c(\textbf{k}))/\left|\partial_{k_0}\mathcal{H}_c\right|$ and the chain-rule identities (see, e.g.,~Ref.~\cite{WeinbergPlasma1962})
\begin{equation}\label{eq:ExplicitBoltzmann}
  \textbf{v}_g =  \frac{\nabla_\textbf{k}  \mathcal{H}}{\partial_{k_0} \mathcal{H} },  \qquad  \nabla_\textbf{X}E(\textbf{k},X)  =  \frac{\nabla_\textbf{X}  \mathcal{H}}{\partial_{k_0} \mathcal{H} },   \qquad  \partial_t E(\textbf{k},X)  =  \frac{\partial_t  \mathcal{H}}{\partial_{k_0} \mathcal{H} }.
\end{equation}
In addition, we used the fact that the denominator $|\partial_{k_0}\mathcal{H}|$ can be written as ${\rm sign}(\partial_{k_0} \mathcal{H}) \partial_{k_0} \mathcal{H}$, with the sign function cancelling on the left- and right-hand sides of Eq.~\eqref{eq:BoltzmannTime}.   

We identify $\textbf{v}_g$ as the group velocity of the mode and $\nabla_\textbf{X} E$ as the standard force term. The term $\partial_t E$ gives a generalisation of the force to time-dependent backgrounds. Note that for an anisotropic medium, the group velocity is, in general, not parallel to $\textbf{k}$ (see, for instance, Ref.~\cite{Befki1966}).

\subsection{Gauge Invariance and Physical Electromagnetic Fields}\label{sec:GaugeInvcePhysFields}

The left-hand side of Eq.~\eqref{eq:BoltzmannTime} is expressed in terms of physical quantities. We would also like to express the collision term on the right-hand side using gauge-invariant objects involving the electromagnetic fields. We begin by noting that the terms $\varepsilon^*_\mu \varepsilon_\nu \Pi^{\lessgtr \, \mu \nu}$ and $\partial_{k_0} \mathcal{H}$ are not separately gauge invariant. The reason for this is that the polarisation vectors $\varepsilon_\mu$ are normalised to unity with $\eta^{\mu \nu} \varepsilon^*_\mu \varepsilon_\nu = 1$, and the covariant normalisation is not conserved under Lorentz transformations. However, the ratio of these two quantities \textit{is} Lorentz invariant, and can therefore be recast in terms of physical fields. To show this, we start by writing
\begin{equation}\label{eq:RatioCollision}
	\frac{ \varepsilon^*_\mu \varepsilon_\nu \Pi^{\lessgtr\, \mu \nu}  }{   \partial_{k_0} \mathcal{H} }  =  \frac{ \mathcal{E}^*_\mu \mathcal{E}_\nu \Pi^{\lessgtr\, \mu \nu}  }{ \mathcal{E}^2 \partial_{k_0} \mathcal{H} } ,
\end{equation}
where now $\mathcal{E}$ is an arbitrarily normalised vector with $\mathcal{E}_\mu \propto \varepsilon_\mu$, whose norm is given by $\mathcal{E}^2 = \mathcal{E}^*_\mu \mathcal{E}_\nu \eta^{\mu \nu}$. Under gauge transformations, the gauge-field $A_\mu$ transforms as
\begin{equation}
A_\mu \rightarrow A_\mu + \partial_\mu \Lambda,
\end{equation}
for some function $\Lambda$. In momentum (Wigner) space, the polarisation vectors $\mathcal{E}$ therefore transform as
\begin{equation}\label{eq:PolarisationTransformation}
	\mathcal{E}_\mu(k) \rightarrow \mathcal{E}_\mu(k)  - i k_\mu \tilde{\Lambda}(k).
\end{equation}
It then follows that, under gauge transformations,
\begin{equation}
	\mathcal{E}^*_\mu \Pi^{\lessgtr \, \mu \nu} \mathcal{E}_\nu \rightarrow \left( \mathcal{E}^*_\mu(k)  + i k_\mu \tilde{\Lambda}^*(k) \right)\Pi^{\lessgtr \, \mu \nu} \left( \mathcal{E}_\nu(k)  - i k_\nu \tilde{\Lambda}(k) \right) = \mathcal{E}^*_\mu \Pi^{\lessgtr \, \mu \nu} \mathcal{E}_\nu,
\end{equation}
where the terms involving contractions  of $k_\mu$ with $\Pi^{\lessgtr \, \mu \nu}$ all vanish on-shell, since
\begin{equation}
	\Pi^{\lessgtr \, \mu \nu} k_\nu = k_\mu \Pi^{\lessgtr \, \mu \nu} =0.
\end{equation}
This establishes that $\mathcal{E}^* \cdot \Pi^{\lessgtr} \cdot \mathcal{E}$ is gauge invariant. Proving the gauge invariance of $\mathcal{E}^2 \partial_{k_0} \mathcal{H}$ is slightly more involved. We give the full details in Appendix \ref{App:GaugeInvce}. 

We now show further how to express $\mathcal{E}^2 \partial_{k_0} \mathcal{H}$ in terms of the energy density of the plasma mode. Specifically, it is straightforward to show, starting from the identity $\mathcal{E}^* \cdot \mathcal{D} \cdot \mathcal{E}  = \mathcal{H} \mathcal{E}^2$, dividing by the photon frequency $\omega$, differentiating with respect to $\partial_{k_0}$ and putting everything on-shell (i.e., taking $k_0 \rightarrow E_\gamma = \omega$), that
\begin{equation}\label{eq:DerivativeD}
\mathcal{E}^*\cdot \partial_{k_0} (\mathcal{D}/k_0) \cdot \mathcal{E} = (\mathcal{E}^2/\omega) \partial_{k_0}\mathcal{H}.
\end{equation}
 As shown in Appendix \ref{App:GaugeInvce}, since the right-hand side of Eq.~\eqref{eq:DerivativeD} is gauge invariant, it follows that the left-hand side can be evaluated in any gauge to derive an expression for $\mathcal{E}^2  \partial_{k_0}\mathcal{H}$.  We choose temporal gauge in which $\mathcal{E}^0 =0$ (and drop the covariant gauge term by taking $\xi \rightarrow \infty$ in Eq.~\eqref{eq:DRInverse}). In this case, the electric field satisfies $E_i = F_{0 i} = \partial_0 A_i = - i \omega \mathcal{E}_i$. We can then write $\Pi^{ij}_{\rm H} = \omega^2 (\delta^{i j} - \epsilon^{ij})$, where $\epsilon^{ij}$ and $\Pi^{ij}_{\rm H}$ are the spatial parts of the dielectric and polarisation tensors, respectively. Using these relations and the explicit form of Eq.~\eqref{eq:DRInverse} allows us to write Eq.~\eqref{eq:DerivativeD} as
\begin{equation}\label{eq:ElectricFieldEquation}
E^*_i \, \partial_\omega \left( \frac{\left| \textbf{k}\right|^2}{\omega} \delta_{i j} - \frac{\textbf{k}_i \textbf{k}_j}{\omega} - \omega \epsilon_{i j}\right) E_j =  \frac{\left|\textbf{E}\right|^2}{\omega} \partial_{k_0} \mathcal{H}.
\end{equation}
 We can then use Faraday's law $\partial_t \textbf{B} =  - \nabla \times \textbf{E} $, which implies $\omega^2 \left|\textbf{B}\right|^2 = \left| \textbf{k}\right|^2 \left|\textbf{E}\right|^2 - (\textbf{k}\cdot \textbf{E})^2$ in momentum space. This can be inserted into Eq.~\eqref{eq:ElectricFieldEquation}, giving
\begin{equation}\label{eq:RM}
   \partial_{k_0} \mathcal{H} =   \omega \, \frac{\left|\textbf{B}\right|^2  +  \textbf{E}^*_i  \partial_\omega (\omega \epsilon_{ i j }) \textbf{E}_j }{\left|\textbf{E}\right|^2}, 
\end{equation}
which, using the temporal gauge identity $\mathcal{E}^2 = -  \left|\textbf{E}\right|^2/\omega^2$, can be written as
\begin{equation}\label{eq:EigenvalueDerivaive}
   \mathcal{E}^2 \partial_{k_0} \mathcal{H}  = -\frac{  \, \left|\textbf{E}\right|^2}{\omega R}.
\end{equation}
Herein, we have defined
\begin{equation}\label{eq:R}
R  = \frac{U_E}{U} = \frac{\left|\textbf{E}\right|^2} {\left|\textbf{B}\right|^2  +  \textbf{E}^*_i  \partial (\omega \epsilon_{ i j }) \textbf{E}_j }
\end{equation}
as the ratio between the electric energy density $U_E$ given by the numerator, and total energy density $U$ in the denominator \cite{Swanson1989,Millar2021}. Relations similar to \eqref{eq:RM}, which relate derivatives of the eigenvalues of the wave operator \eqref{eq:DRInverse} to the ratio of stored energies, have been discussed at length in Ref.~\cite{Melrose1981} and Sec.~2.3.6 of Ref.~\cite{MelroseBookI_QPD2008}.

Finally, since $\mathcal{E}^* \cdot \Pi^{\lessgtr} \cdot \mathcal{E}$ is also gauge invariant, it too can be evaluated in temporal gauge, to give
\begin{equation}\label{eq:PiEEPhysical}
	\mathcal{E}^*  \cdot \Pi^\lessgtr \cdot \mathcal{E} = \left| \textbf{E}\right|^2 \hat{\boldsymbol{\varepsilon}}^*_i \hat{\boldsymbol{\varepsilon}}_j  \Pi^{\lessgtr\, ij}.
\end{equation}
Here, we have again used $\mathcal{E}^*_i \mathcal{E}_i = \left|\textbf{E}\right|^2/\omega^2$ and defined
\begin{equation}
	\hat{\boldsymbol{\varepsilon}} = \textbf{E}/\left|\textbf{E} \right|,
\end{equation}
as the unit 3-polarisation of the electric field for the given eigenmode. Putting all these steps together, we can insert Eqs.~\eqref{eq:EigenvalueDerivaive} and \eqref{eq:PiEEPhysical} into Eq.~\eqref{eq:RatioCollision} to get
\begin{equation}
\frac{ \varepsilon^*_\mu \varepsilon_\nu \Pi^{\lessgtr\, \mu \nu}  }{  \partial_{k_0} \mathcal{H} }  =  - \frac{ U_E}{\omega U} \hat{\boldsymbol{\varepsilon}}^*_i \hat{\boldsymbol{\varepsilon}}_j  \Pi^{<\, ij}.
\end{equation}
This can then be inserted into the right-hand side of Eq.~\eqref{eq:BoltzmannTime} to yield a form of the Boltzmann equation in terms of electromagnetic fields
\begin{equation}\label{eq:BoltzmannEMFields}
\partial_t f_\gamma + \textbf{v}_g \cdot \nabla_\textbf{x} f_\gamma - \nabla_\textbf{x} E_\gamma \cdot \nabla_\textbf{k} f_\gamma - \partial_t E_\gamma \partial_{E} f_\gamma = -  \frac{ U_E}{ \omega U} \hat{\boldsymbol{\varepsilon}}^*_i \hat{\boldsymbol{\varepsilon}}_j \Big( (1+ f_\gamma) \Pi^{<\, ij}  - \Pi^{>\, i j} f_\gamma \Big).
\end{equation}
Note that in the equation above, and henceforth in the main text, we shall simply write lower-case $\textbf{x}$ instead of $\textbf{X}$ for the central coordinate - this is partly for ease of notation and to make its interpretation as a position phase-space coordinate more readily apparent. The appearance of the stored energy in the plasma, as captured by the factor $R = U_E/U$ defined in Eq.~\eqref{eq:R}, has also been identified in scattering processes in plasmas \cite{Melrose1981} and was discussed in Ref.~\cite{Millar2021}.

\subsection{Axion Source Term }

Having derived a consistent set of transport equations that determine the evolution of the photon distribution functions, we now turn our attention to isolating the photon production process via axions. To this end, we assume that the axions and photons are not in equilibrium with each other. We also neglect the emission of photons directly from the plasma, and focus on the regime where the mean free path of photons is much larger than the spatial extent of the plasma, such that absorption processes can be neglected. Formally, this corresponds to the decomposition of the photon Wightman self-energy into
\begin{equation}
\Pi^{\lessgtr} = \Pi_{\rm pl}^{\lessgtr} + \Pi_{\rm ax}^{\lessgtr},
\end{equation}
where $\Pi_{\rm pl}^{\lessgtr}$ and $\Pi_{\rm ax}^{\lessgtr}$ correspond to the production/absorption of photons due to plasma processes and the axion background, respectively. The approximation amounts to isolating the contribution to photon production from $\Pi_{\rm ax}^{\lessgtr}$ and neglecting any effects arising form $\Pi_{\rm pl}^{\lessgtr}$. 

Working then in the limit that $f_{\gamma}\ll 1$, we can take the Born approximation in the collision terms and only the gain term remains on the right-hand side of the Boltzmann equation. In this regime, putting Eq.~\eqref{Eq:BoltzmannFull} on-shell gives the following equation for the photon distribution functions:
\begin{equation}\label{eq:TransportBornApprox}
    \partial_k \mathcal{H} \partial_x f_{\gamma}	- \partial_x \mathcal{H} \partial_k f_\gamma  =  \varepsilon^*_\mu \varepsilon_\nu   \Pi_{\rm ax}^{< \mu \nu}.
\end{equation}
We reiterate that the polarisation eigenlabel $c$ on $f^c_\gamma$, $\varepsilon^c$ and $\mathcal{H}^c$ has been suppressed; it is understood that we deal with production of a particular photon eigenmode, whose distribution function we label simply $f_\gamma$. 

\begin{figure}
	\centering
	\includegraphics{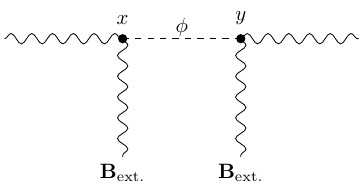}
	\caption{Contribution to the photon Wightman self-energy  $\Pi_{\rm ax}^<(x,y)$ due to axion interactions. The optical cut through the axion line gives the standard inverse Primakoff process $a \rightarrow \gamma$.}
	\label{fig:SelfEnergy}
\end{figure}

We recall that the object $\mathcal{H}=\mathcal{H}(k,x)$ is the Hamiltonian of the mode, with $k_\mu$ and $x^\mu$ giving the 4-momentum and coordinates of the photon, respectively. Setting $\mathcal{H}(k,x) =0$ gives the dispersion relation for that mode. For example, in an unmagnetised cold plasma, one has $\mathcal{H}= k_\mu k^\mu - \omega_p^2$. The object $\Pi_{\rm ax}^{ < \mu \nu}(k,x)$ is the Wigner transform of the photon Wightman self-energy, and corresponds to production due to axions. It is defined by
\begin{equation}\label{eq:Piax}
	\Pi^{< \mu \nu}_{\rm ax}(x,x') =\braket{j^\nu_\phi(x')j^\mu_\phi(x) } .
\end{equation}
Here, $j_\phi^{\mu}$ is the effective current arising from the axion-photon interaction
\begin{equation}\label{eq:Lint}
	\mathcal{L}_{\rm int} = - \frac{g_{a \gamma \gamma}}{4} \phi F_{\mu \nu} \tilde{F}^{\mu \nu} =   A_\mu j^\mu_\phi, \qquad j_\phi^\mu = g_{a \gamma \gamma} \partial_\nu \phi  \tilde{F}^{\mu \nu }_{\rm ext} \, ,
\end{equation}
where $\tilde{F}^{\mu \nu }_{\rm ext}$ is the dual external field strength tensor, originating, for instance, from a background magnetic field. The form of the production term on the right-hand side of Eq.~\eqref{eq:TransportBornApprox} has the form of the standard photon-production term for gauge theories \cite{Caron-Huot:2006pee,Gervais:2012wd,Bellac:2011kqa,Kapusta:2006pm}, except that here it applies to more general polarisations relevant for arbitrary inhomogeneous and anisotropic media. It is straightforward to show that, to leading order in $g_{a \gamma \gamma}$ (see Appendix \ref{sec:collision}), the Wightman self-energy contribution from axions is given by
\begin{equation}\label{eq:SelfEnergy}
	\Pi^{< \, \mu \nu}_{\rm ax}(k,X) = g_{a \gamma \gamma}^2 k_\rho k_\sigma \tilde{F}_{\rm ext}^{\mu \rho} \tilde{F}_{\rm ext}^{\nu \sigma}  2 \pi \delta(k^2 - m_\phi^2) f_\phi(k,X) \ ,
\end{equation}
where $f_\phi(k,x)$ is the phase-space density of axions. The result \eqref{eq:SelfEnergy} is generated from the diagram in Fig.~\ref{fig:SelfEnergy}.

Furthermore, we should remember that the $k_0$ appearing in Eq.~\eqref{eq:SelfEnergy} is on-shell with respect to the photon dispersion relation, i.e., $k_0 = E_\gamma(\textbf{k},x)$. Since the axion energy satisfies $E_\phi(\textbf{k})^2 = \left|\textbf{k}\right|^2 + m_\phi^2$, it follows that the delta function can be written as
\begin{equation}\label{eq:EnergySplit}
	\delta(k^2 - m_\phi^2) = \delta \left(E_\gamma(\textbf{k},x)^2 - E_\phi(\textbf{k})^2 \right).
\end{equation}
By inserting Eq.~\eqref{eq:EnergySplit} into Eq.~\eqref{eq:SelfEnergy} and then placing this into Eq.~\eqref{eq:TransportBornApprox}, we arrive at the following form of the transport equation for photons:
\begin{equation}\label{eq:BoltzmannExplicit}
   \partial_k \mathcal{H} \partial_x f_\gamma	- \partial_x \mathcal{H} \partial_k f_\gamma  =   g_{a \gamma \gamma}^2 \big| k \cdot \tilde{F}_{\rm ext} \cdot \varepsilon \big|^2 2 \pi \delta \left(E_\gamma ( \textbf{k},x)^2 - E_\phi(\textbf{k})^2 \right) f_\phi \, ,
\end{equation}
where the dots on the right-hand side denote contraction of Lorentz indices. Note that production is on a mode-by-mode basis, i.e., this equation describes conversion from axions into a given photon eigenmode. We emphasise again that the polarisation 4-vector $\varepsilon$ is computed \textit{in-medium} and is not simply the vacuum polarisation vector.

In summary, we see formally that, by solving Eq.~\eqref{eq:BoltzmannExplicit} for $f_\gamma$, we directly obtain the asymptotic distribution of photons as they escape from magnetised plasma regions and become transversely polarised photons in vacuum. This recasts the problem as one of radiative transfer.

\section{Photon Production and Propagation}\label{sec:PhotonProuction}
The result \eqref{eq:BoltzmannExplicit} gives the governing equation for the phase-space density of photons. By solving this 8-dimensional phase-space equation, one could in principle completely reconstruct the phase-space density of photons. However, it is also natural to use a ray-tracing prescription in which we track trajectories of individual particles. This is equivalent to solving this equation through a method of characteristics. Indeed, ray-tracing approaches have been used extensively \cite{Leroy:2019ghm,Witte:2021arp,Battye:2021xvt,FosterSETI2022,Battye:2023oac,Battye2022} to derive astrophysical signals of axions. For this reason, it is more useful to solve these equations for $f_\gamma$ computed along the worldline of photons, whose trajectories can then be used to construct observables, such as the radio flux from an astrophysical environment.

\begin{figure}
	\centering
	\includegraphics{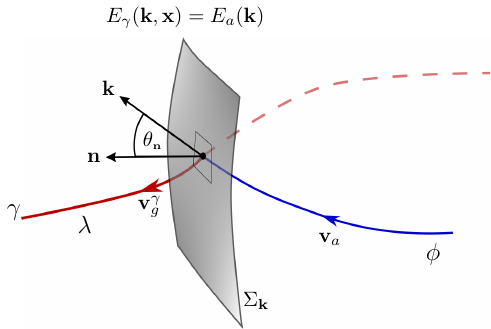}
	\caption{Worldline method for computing axion-photon conversion. The phase-space density $f_\gamma$ is integrated along the photon worldline (red) with affine parameter $\lambda$. The axion worldline is shown in blue. Production occurs at the point $E_\gamma\big(\textbf{k}(\lambda),\textbf{x}(\lambda)\big) = E_a\big(\textbf{k}(\lambda)\big)$ on the kinematic surface $\Sigma_\textbf{k}$ with unit normal $\textbf{n} \propto \nabla E_\gamma$. The axion and photon 4-momenta are equal at the resonance $k_a^\mu = k^\mu_\gamma$, but their group velocities, $\textbf{v}_g^\gamma$ and $\textbf{v}_a \propto \textbf{k}$, which are tangent to worldlines, are different, since in an anisotropic medium, the photon 3-momentum and group velocity are not parallel \cite{Befki1966}.}
	\label{fig:Rays}
\end{figure}

With this in mind, we construct solutions using a method of characteristics, which consists of identifying those integral curves $(x^\mu (\lambda),k_\mu(\lambda))$ in phase space that satisfy
\begin{equation}\label{eq:HamiltonEqs}
   \frac{{\rm d} x^\mu (\lambda)}{{\rm d} \lambda}   = \frac{\partial  \mathcal{H}}{\partial k_\mu}, \qquad \frac{{\rm d} k_\mu (\lambda)}{{\rm d} \lambda}   =-  \frac{\partial  \mathcal{H}}{\partial x^\mu} \, ,
\end{equation}
where $\lambda$ is a worldline parameter. From the chain rule, it follows that along these curves, Eq.~\eqref{eq:BoltzmannExplicit} reads
\begin{equation}\label{eq:RadiativeTransfer}
   \frac{{\rm d} f_\gamma(k(\lambda),x(\lambda))}{{\rm d} \lambda}  =  g_{a \gamma \gamma}^2 \big|k \cdot \tilde{F}_{\rm ext} \cdot \varepsilon \big|^2  2 \pi \delta \left(E_\gamma ( \lambda)^2 - E_\phi(\lambda)^2 \right) f_\phi.
\end{equation}
By solving this equation and propagating the solutions out to infinity, one immediately obtains the asymptotic distribution of photons measured by an external observer. This situation is illustrated in Fig.~\ref{fig:Rays}. Note also that $E_{\gamma, \phi} ( \lambda)$ is a shorthand for $E_{\gamma, \phi}\left(\textbf{k}(\lambda),x(\lambda)\right)$ and that the axion energy function is evaluated along the \textit{photon} worldline. 
Working, as before, in temporal gauge with $\varepsilon^0 =0$ and integrating Eq.~\eqref{eq:RadiativeTransfer} gives
\begin{equation} \label{eq:fRes}
	f_\gamma(\lambda) = \frac{ \pi g_{a \gamma \gamma}^2 E_\gamma^2 \big| \textbf{B}_{\rm ext}\cdot \boldsymbol{\varepsilon}\big|^2 }{ E_\gamma \big| E_\gamma'(\lambda_c) - E_\phi'(\lambda_c) \big|} f_\phi(\textbf{k}(\lambda_c),x(\lambda_c))
\end{equation}
for $\lambda > \lambda_c$, where $\lambda_c$ is the value of the worldline parameter for which the energies satisfy
\begin{equation}
E_\gamma ( \lambda_c)^2 - E_\phi(\lambda_c)^2 = 0.
\end{equation}
All the quantities on the right-hand side of Eq.~\eqref{eq:fRes} are understood to be evaluated at the resonance $\lambda = \lambda_c$. Primes in the denominator of Eq.~\eqref{eq:fRes} denote differentiation with respect to $\lambda$.  Now, for a stationary background, $E_\gamma$ is conserved, so that $E_\gamma'(\lambda) =0$. By using $E^2_\phi(\lambda) = \textbf{k}(\lambda)\cdot \textbf{k}(\lambda) + m_\phi^2$, we then have
\begin{equation}\label{eq:dEdlambda}
	E_\phi'(\lambda) =  \frac{\textbf{k}(\lambda)\cdot \textbf{k}'(\lambda)}{E_\phi} = - \frac{\textbf{k}(\lambda)\cdot \partial_\textbf{x}\mathcal{H} }{E_\phi} = - \partial_{k_0} \mathcal{H} \, \textbf{v}_p \cdot \nabla_\textbf{x} E ,
\end{equation}
where, in the final equality, we used the chain-rule identity \eqref{eq:ExplicitBoltzmann}. We emphasise again that $\textbf{k}(\lambda)$ is the 3-momentum along the \textit{photon} worldline. Defining a ``conversion probability'' $P_{a \gamma}$ via
\begin{equation}
	f_\gamma = P_{a \gamma } f_\phi,
\end{equation}
we can read off
\begin{equation}\label{eq:ProbabilityCovariant}
	P_{a \gamma  } =   \frac{  \pi g_{a \gamma \gamma}^2 \big| k \cdot \tilde{F}_{\rm ext} \cdot \varepsilon \big|^2	}{ E_\gamma \partial_{k_0} \mathcal{H} \left| \textbf{v}_p \cdot \nabla_\textbf{x} E_\gamma(\textbf{k},\textbf{x})\right|} ,
\end{equation}
where $\textbf{v}_p = \textbf{k}/E_\gamma$ is the phase velocity of the photon, equal to the phase velocity of the axion at the resonance by energy-momentum conservation.  We can use results from the previous section (especially Eq.~\eqref{eq:RM}) to re-express this as
\begin{equation}\label{eq:ProbabilityPhysical}
	P_{a \gamma  } =   \frac{  \pi g_{a \gamma \gamma}^2 \big|\hat{\boldsymbol{\varepsilon}}\cdot \textbf{B}_{\rm ext}\big|^2	}{  \left| \textbf{v}_p \cdot \nabla_\textbf{x} E_\gamma(\textbf{k},\textbf{x})\right|}  \frac{U_E}{U},
\end{equation}
where $\hat{\boldsymbol{\varepsilon}}$ is the unit polarisation 3-vector of the electric field and $\textbf{B}_{\rm ext}$ is the external magnetic field.  It is important to remark that this equation is valid for any quasi-stationary medium, so long as the energy (i.e., dispersion relation) and polarisation vectors are known. Notice in particular that the gradient term in the denominator is projected along the axion worldline, which is in the direction of $\textbf{k}$, rather than the photon worldline, which is in the direction of $\textbf{v}_g$. In an anisotropic medium, $\textbf{v}_g$ and $\textbf{k}$ are in general not parallel \cite{Befki1966}.

The arguments laid out in this and previous sections have provided a rigorous means both for deriving and solving transport equations for photons to compute the production due to axions. Nonetheless, it is interesting to note that the same equation \eqref{eq:ProbabilityPhysical} emerges from more heuristic arguments involving naive second quantisation of plasma modes. To see this, one can take the classical modes in the medium and apply second quantisation to these quasiparticles so that the gauge field can be written as  \cite{Melorse1983II}
\begin{equation}\label{eq:2ndQuant}
A^\mu(x) = \sum_c \int \frac{{\rm d}^3 \textbf{k}}{(2 \pi)^3 \sqrt{2 E_\gamma}} \left[ \varepsilon^\mu_c c^c_\textbf{k} e^{ - i k\cdot x} + \varepsilon^{* \, \mu}_c c^{c \, \dagger}_\textbf{k} e^{i k\cdot x} \right],
\end{equation}
where $c^{c \, \dagger}_\textbf{k}$ and $c^c_\textbf{k}$ are creation and annihilation operators for the plasma eigenmode $c$. We emphasise again that these create collective plasma excitations, not vacuum photons, such that $\varepsilon_c$ are the (local) in-medium polarisation vectors of the inhomogeneous plasma. In this case, one can simply ``write down'' a Boltzmann equation
\begin{equation}
   \left\{ f_\gamma, \mathcal{H} \right\} =
	\int \frac{{\rm d}^3 \textbf{k}_\phi}{(2\pi)^3 2 E^\phi_\textbf{k}} \left| \mathcal{M}_{\phi \rightarrow \gamma}\right|^2 (2 \pi)^4 \delta^{(4)}(k - k_\phi) f_\phi,
\end{equation}
where the right-hand side is a collision integral containing the matrix element for axion-photon conversion into a particular photon mode. The latter can be read off from Eqs.~\eqref{eq:Lint} and \eqref{eq:2ndQuant}, giving
\begin{equation}
	\left| \mathcal{M}_{\phi \rightarrow \gamma}\right|^2  =\left|\bra{\gamma} \mathcal{L}_{\rm int} \ket{\phi}\right|^2 =g_{a \gamma \gamma}^2 \big| k \cdot \tilde{F}_{\rm ext} \cdot \varepsilon\big|^2.
\end{equation}
  Upon performing the 3-momentum integration in the collision integral, we get
\begin{equation}
   \left\{ f, \mathcal{H} \right\} =
	2 \pi   g_{a \gamma \gamma}^2 \big|k \cdot \tilde{F}_{\rm ext} \cdot \varepsilon \big|^2 \frac{ \delta\left(E_\gamma(x, \textbf{k}) - E_\phi(\textbf{k})\right)}{2 E_\gamma(x, \textbf{k})}  f_\phi.
\end{equation}
We can again project this equation onto worldlines, as was done previously, to arrive at
\begin{align}
   \frac{{\rm d} f_\gamma}{{\rm d} \lambda}  & =  \pi   g_{a \gamma \gamma}^2 \big| k \cdot \tilde{F}_{\rm ext} \cdot \varepsilon\big|^2 \frac{ \delta\left(E_\gamma(\lambda) - E_\phi(\lambda) \right)}{ E_\gamma}  f_\phi.
\end{align}
Integrating this equation over $\lambda$ and using Eq.~\eqref{eq:dEdlambda}, we arrive again at Eq.~\eqref{eq:ProbabilityCovariant}, justifying our conversion probability in terms of more familiar quantum field theory arguments involving Boltzmann equations and collision integrals.

\subsection{Total Power}\label{sec:ContinuitySec}

It may look like the denominator in the conversion probability (Eqs.~\eqref{eq:ProbabilityCovariant} and \eqref{eq:ProbabilityPhysical}) gives uncontrolled divergences when $\textbf{v}_p$ is perpendicular to $\nabla_\textbf{x} E_\gamma$. However, when integrated over phase space, this would-be divergence is cancelled by the phase-space measure, as we now explain.

To show this, we first return to the Boltzmann equation \eqref{eq:BoltzmannExplicit}, which can be multiplied by $k_0$ to arrive at
\begin{equation}
\Big\{ \mathcal{H},
f_\gamma  \Big\} k_0 \delta(\mathcal{H})=g_{a \gamma \gamma}^2 \big|k \cdot \tilde{F}_{\rm ext} \cdot \varepsilon\big|^2 \delta \left(k_0^2 - E_\phi(\textbf{k})^2 \right)f_\phi \, k_0  \delta(\mathcal{H}).
\end{equation}
Note that the left-hand side can be written as
\begin{align}
\Big\{ \mathcal{H},
f  \Big\} k_0 \delta(\mathcal{H}) & = k_0 \Big\{ \mathcal{H},
f  \delta(\mathcal{H}) \Big\} \nonumber \\
& = k_0 \partial_k \mathcal{H} \partial_x \left(f \delta(\mathcal{H})\right) - k_0 \partial_x \mathcal{H} \partial_k \left(f \delta(\mathcal{H})\right) \nonumber \\
& = k_0 \partial_x \left[ \partial_k \mathcal{H} f \delta(\mathcal{H})\right] - k_0 \partial_k \left[  \partial_x \mathcal{H} f \delta(\mathcal{H}) \right].
\end{align}
Next, we can integrate this equation over a \textit{finite} spatial 3-volume $\mathcal{V}$, whose bounding surface has area element ${\rm d}A$, and a region in 4-momentum space $\int {\rm d}^4 k$ running over all momenta. Upon performing these integrations, we arrive at
\begin{equation}\label{eq:EnergyConservation}
	\frac{{\rm d}}{{\rm d}t} \int {\rm d}\mathcal{V} \int {\rm d}^3\textbf{k} \, \omega f_\gamma + \int {\rm d}^3 \textbf{k} \int {\rm d}\textbf{A} \cdot \textbf{v}_g \, \omega f_\gamma +  \int {\rm d}^3 \textbf{k} \int {\rm d} \mathcal{V} \, \partial_t E_\gamma f_\gamma = \int {\rm d}\mathcal{V} \, Q,
\end{equation}
where we have used $\partial_\textbf{k}  \mathcal{H} /\partial_{k_0} \mathcal{H}  = \textbf{v}_g$  and $\partial_\textbf{x} \mathcal{H}/\partial_{k_0} \mathcal{H} = \nabla_\textbf{x} E_\gamma$. To derive the second term, we applied the divergence theorem with respect to the spatial integral. We also discarded the surface terms arising from the ${\rm d}^4 k$ integration of $k_0 \partial_k \left[  \partial_x \mathcal{H}_c  f^c \delta(\mathcal{H}_c) \right]$. The source term $Q$ is defined as
\begin{equation}
	Q = \int {\rm d}^3\textbf{k} \;  \omega g_{a \gamma \gamma}^2 (k \cdot \tilde{F}_{\rm ext} \cdot \varepsilon^c)^2 2 \pi \delta \left(E_\gamma ( \textbf{k},x)^2 - E_\phi(\textbf{k})^2 \right) \frac{1}{\partial_{k_0}  \mathcal{H}} f_\phi .
\end{equation}
The integral identity in Eq.~\eqref{eq:EnergyConservation} is, of course, nothing more than a continuity equation for the photon flux. The first term gives the rate of change of the total energy in the volume $\mathcal{V}$, which changes due to flux through the boundary (left-hand side, second term), energy shifts in photons due to a non-stationary background (left-hand side, third term), or due to production (right-hand side).

Next, we can make use of the integral identity 
\begin{equation}
    \int {\rm d}^n \textbf{x} \; \delta(G(\textbf{x})) = \int_{G^{-1}(0)} {\rm d}\Sigma  \,\frac{1}{\left| \nabla G \right|},
\end{equation}
where ${\rm d}\Sigma$ is the area element on the surface defined by $G(\textbf{x}) =0$. Applying this to the spatial integration on the right-hand side of Eq.~\eqref{eq:EnergyConservation}, by treating $\delta \left(E_\gamma ( \textbf{k},\textbf{x})^2 - E_\phi(\textbf{k})^2 \right)$ as a function of $\textbf{x}$ for fixed $\textbf{k}$, we arrive at
\begin{equation}\label{eq:QIntegral}
	\int {\rm d}\mathcal{V} \, Q = \int {\rm d}^3 \textbf{k} \int \, {\rm d} \Sigma_\textbf{k} \,  \omega \, \,  \frac{ \pi g_{a \gamma \gamma}^2 \big| k \cdot \tilde{F}_{\rm ext} \cdot \varepsilon\big|^2}{ E_\gamma \partial_{k_0} \mathcal{H} \left| \nabla_\textbf{x} E_\gamma \right| } \, f_\phi,
\end{equation}
where $\Sigma_\textbf{k}$ is the spatial surface on which $E_\gamma(\textbf{k},\textbf{x}) = E_\phi(\textbf{k})$. We can leave the integrand in Eq.~\eqref{eq:QIntegral} unchanged by multiplying the numerator and denominator by the same factor $ v_p \cos \theta_\textbf{n}$, where $v_p$  is the magnitude of the phase velocity and $\theta_\textbf{n}$ is the angle between the phase velocity and the surface normal of $\Sigma_\textbf{k}$ denoted by $\textbf{n}$. By doing this, we have
\begin{equation}
	\frac{{\rm d} \Sigma_\textbf{k}}{\left| \nabla_\textbf{x} E_\gamma \right| } =  \frac{{\rm d} \Sigma_\textbf{k} \cos \theta_\textbf{n} v_p }{\cos \theta_\textbf{n}  v_p\left| \nabla_\textbf{x} E_\gamma \right| }  =  \frac{{\rm d} \boldsymbol{\Sigma}_\textbf{k} \cdot \textbf{v}_p }{\left| \textbf{v}\cdot \nabla_\textbf{x} E_\gamma \right| },
\end{equation}
where ${\rm d}\boldsymbol{\Sigma}_\textbf{k} = \textbf{n} {\rm d} \Sigma_\textbf{k}$ is the directed surface element, so that $\textbf{n}\cdot \hat{\textbf{v}}_p = \cos \theta_\textbf{n}$. Hence, we have
\begin{align}
	\int {\rm d}\mathcal{V} \, Q & = \int {\rm d}^3 \textbf{k} \int \, {\rm d} \boldsymbol{\Sigma}_\textbf{k}\cdot \textbf{v}_p \, \omega \,  \frac{ \pi g_{a \gamma \gamma}^2 \big| k \cdot \tilde{F}_{\rm ext} \cdot \varepsilon \big|^2}{ E_\gamma \partial_{k_0} \mathcal{H} \left| \textbf{v}_p \cdot \nabla_\textbf{x} E_\gamma \right| } \, f_\phi \nonumber \\
	&=  \int {\rm d}^3 \textbf{k} \int \, {\rm d} \boldsymbol{\Sigma}_\textbf{k}\cdot \textbf{v}_p \omega \, P_{a \gamma} f_\phi .
\end{align}
Inserting this into Eq.~\eqref{eq:EnergyConservation} gives (for a stationary Hamiltonian)
\begin{equation}\label{eq:ContinuityFinal}
	\frac{{\rm d}}{{\rm d}t} \int {\rm d}\mathcal{V} \int {\rm d}^3\textbf{k} \, \omega \, f_\gamma + \int {\rm d}^3 \textbf{k} \int {\rm d}\textbf{A} \cdot \textbf{v}_g \omega f_\gamma = \int {\rm d}^3 \textbf{k} \int {\rm d}  \boldsymbol{\Sigma}_\textbf{k}\cdot \textbf{v}_p \omega P_{a \gamma} f_\phi.
\end{equation}

From the right-hand side of Eq.~\eqref{eq:ContinuityFinal}, we immediately see that the would-be divergence in the probability, arising when $\textbf{v}_p$ is perpendicular to $\nabla_\textbf{x} E_\gamma$, is cancelled by the flux-like projection ${\rm d}\boldsymbol{\Sigma}_\textbf{k} \cdot \textbf{v}_p$ in the phase-space measure. Note, however, that the limit $\nabla_\textbf{x} E_\gamma \rightarrow 0$ \textit{does} represent a divergence, corresponding to strongly adiabatic conversion. This case is more complicated and will be discussed in the conclusions.

\section{Magnetised Plasmas}\label{sec:MagnetisedPlasmas}

We now come to one of the main physical situations in which this conversion can be active:\ magnetised astrophysical plasmas, as are encountered in the magnetospheres of neutron stars.  

\subsection{Weakly Magnetised Plasmas}
In a weakly magnetised, cold plasma, $\omega_c \ll \omega, \omega_p$,  where $\omega = E_\gamma$ is the photon-frequency (in natural units), $\omega_c =e \left|\textbf{B}\right|/m_e$ is the cyclotron frequency, and $\omega_p = \sqrt{e^2 n_e/m_e}$ is the plasma-frequency with $n_e$ being the electron number density. In that limit, the photon dispersion relation becomes
\begin{equation}
E_\gamma^2 =\left| \textbf{k}\right|^2 + \omega_p^2.
\end{equation}
In addition, $\textbf{v}_p\cdot \nabla E_\gamma= (\omega_p/E_\gamma) \textbf{v}_p \cdot \nabla \omega_p$, and the polarisation vectors are transverse to the direction of propagation, so that $ (\textbf{B}_{\rm ext}\cdot \boldsymbol{\varepsilon})^2 = \sin^2 \theta |\textbf{B}_{\rm ext}|^2$, where $\theta$ is the angle between $\textbf{k}$ and $\textbf{B}_{\rm ext}$. The ratio of the total and electric energy in a weakly magnetised plasma turns out to be $U/U_E = 2$. All together, using Eq.~\eqref{eq:ProbabilityPhysical}, this yields
\begin{equation}\label{eq:IsoProb}
	P^{\rm iso}_{a \gamma \gamma } = \frac{\pi}{2}  \frac{ g_{a \gamma \gamma}^2 \sin^2 \theta |\textbf{B}_{\rm ext}|^2 }{\left|\textbf{v}_p\cdot \nabla \omega_p\right|}  \frac{E_\gamma}{\omega_p}.
\end{equation}
This reproduces the 1D formula for propagation perpendicular to $\textbf{B}$ ($\theta = \pi/2$) in the non-relativistic limit $E_\gamma/\omega_p \simeq 1$, as derived in Refs.~\cite{Battye_2020,Leroy:2019ghm}.

\subsection{Strongly Magnetised Plasmas}
For a strongly magnetised plasma, we can use the relation between the permittivity and the polarisation tensor
\begin{equation}\label{eq:3polarisationTensor}
	\Pi^{ij}  = \omega^2 (\delta^{ij} - \epsilon^{ij}) .
\end{equation}
We choose coordinates in which $\textbf{k}=(0,0,k)$ points in the z-direction, and $\textbf{B}_{\rm ext}$ lies in the $y$-$z$ plane, with $\textbf{B}_{\rm ext} = \left| \textbf{B}_{\rm ext} \right|( - \sin \theta \hat{\textbf{y}} +  \cos \theta \hat{\textbf{z}})$, so that $\theta$ is the angle between $\textbf{k}$ and $\textbf{B}$.  For a strongly magnetised plasma, neglecting relativistic effects\footnote{The generalisation to the relativistic limit is straightforward, see, e.g., Ref.~\cite{Witte:2021arp}.}, we find the permittivity 3-tensor is given by \cite{gurevich_beskin_istomin_1993}
\begin{equation}\label{eq:Permittivity}
\epsilon =
\begin{pmatrix}
	1 & 0 & 0 \\
	0 & 1 - \frac{{\omega_p}^2}{{\omega}^2} \sin^2 \theta & \frac{{\omega_p}^2}{{\omega}^2} \cos \theta \sin \theta \\
	0 & \frac{{\omega_p}^2}{{\omega}^2} \cos \theta \sin \theta & 1 - \frac{{\omega_p}^2}{{\omega}^2} \cos^2 \theta
\end{pmatrix}
.
\end{equation}
From this, we can read off, using the equation for the electric field
inferred from Eqs.~\eqref{eq:OrthonormalBasis}, the following dispersion relations:
\begin{align}
 &k_0^2 - |\textbf{k}|^2 =0, \\  
\nonumber \\
& k_0^2 - \frac{1}{2} \left( |\textbf{k}|^2 + \omega_p^2 + \sqrt{|\textbf{k}|^4  + \omega_p^4 + 2\omega_p^2 |\textbf{k}|^2(1 - 2 \cos^2 \theta)} \right)=0, \\
  \nonumber \\
 & k_0^2 - \frac{1}{2} \left( |\textbf{k}|^2 + \omega_p^2 - \sqrt{|\textbf{k}|^4  + \omega_p^4 + 2\omega_p^2 |\textbf{k}|^2(1 - 2 \cos^2 \theta)} \right) =0 ,
\end{align}
which follow by setting the $\mathcal{H}_c =0$.
These are the magnetosonic-t, Langmuir-O (LO) and  Alfv{\'e}n modes, respectively. To leading order in $g_{a \gamma \gamma}$, the LO mode is the only one which can both be produced on resonance and propagate out of the magnetosphere as $\omega_{\rm p} \rightarrow 0$. The LO mode has electric polarisation unit 3-vector
 \begin{equation}
	\boldsymbol{\varepsilon}_{\rm LO}=\frac{-1}{\sqrt{1+\frac{\omega_p^4\cos^2\theta\sin^2\theta}{(\omega^2-\omega_p^2\cos^2\theta)^2}}}\left({\bf \hat y}-\frac{\omega_p^2\cos\theta\sin\theta}{\omega^2-\omega_p^2\cos^2\theta}{\bf \hat z}\right ) \label{eq:LO}.
\end{equation}
It is also straightforward to show that for the LO mode, $U/U_E = 2$. We verified this by explicit calculation using Eqs.~\eqref{eq:LO} and \eqref{eq:R}. This can also be seen by noting that, since the permittivity takes the form $\epsilon  = I  - A/\omega^2$ for some matrix A, it follows that $\partial_\omega (\omega \epsilon) = I + A/\omega = 2 I - \epsilon $, so that $U = \left|\textbf{B}\right|^2 + \textbf{E}^*\cdot \partial_\omega(\omega \epsilon)\cdot \textbf{E} = \left|\textbf{B}\right|^2 + 2 \left|\textbf{E}\right|^2 - \textbf{E}^*\cdot \epsilon \cdot \textbf{E}$. Again, using Faraday's law to write $\left|\textbf{B}\right|^2 = [ \left| \textbf{k}\right|^2 \left|\textbf{E}\right|^2 - (\textbf{k}\cdot \textbf{E})^2]/\omega^2$,  we have that $U = \omega^2 E^*_i (|\textbf{k}|^2 \delta_{ij}- k_i k_j - \omega^2 \epsilon_{i j})E_j + 2 E^*_i E_i$, but the first term vanishes due to the equations of motion, leaving $U = 2 \left|\textbf{E}\right|^2 = 2 U_E$, as required.

From Eqs.~\eqref{eq:ProbabilityPhysical} and \eqref{eq:LO}, we therefore read off\\
\begin{equation}\label{eq:PStrongB}
	P_{a \gamma } = \frac{\pi}{2} \frac{ g_{a \gamma \gamma}^2\left| \textbf{B}_{\rm ext} \right|^2E_\gamma ^4 \sin ^2 \theta }{\cos ^2 \theta
   \, \omega _p^2 \left(\omega _p^2-2 E_\gamma ^2\right)+E_\gamma ^4} 	\frac{1}{\left| \textbf{v}_p \cdot \nabla_\textbf{x} E_\gamma  \right|} .
\end{equation}
This result reproduces the isotropic result \eqref{eq:IsoProb} in the limit $\theta \rightarrow \pi/2$ in which the LO mode becomes ordinary and behaves as though in an isotropic plasma. Equation \eqref{eq:PStrongB} is the central result of our paper, providing the conversion probability for axions to photons in a strongly magnetised plasma. Note that we have also verified, via an alternative and more explicit calculation, the correctness of the form of Eq.~\eqref{eq:PStrongB}. This we did in covariant gauge using the covariant form \eqref{eq:ProbabilityCovariant} and deriving an explicit form of the eigenvalue derivative $\partial_{k_0} \mathcal{H}$. This calculation is given in the Appendix \ref{Appendix:CovariantGauge}.  

We also state here the explicit form of the energy gradient, which corresponds to taking spatial derivatives while keeping $\textbf{k}$ constant. This gives
\begin{equation}
\nabla_\textbf{x} E_\gamma (\textbf{k},x) =\frac{\omega  \omega _p \left( \cos \theta  \omega _p \left(\omega _p^2-\omega ^2\right) \nabla_\textbf{x} \cos \theta  +\omega ^2  \sin ^2\theta
   \nabla_\textbf{x} \omega_p\right)}{\cos ^2\theta  \left(\omega _p^4-2 \omega ^2 \omega _p^2\right)+\omega ^4},
\end{equation}
where
\begin{align}
\nabla_\textbf{x} \cos \theta &= \nabla_\textbf{x}\left(\frac{\textbf{k}\cdot \textbf{B}}{{\left|\textbf{k}\right| \left|\textbf{B}_{\rm ext} \right|}}\right)  \nonumber \\
& = \frac{k_i  \nabla_\textbf{x} B^i}{\left|\textbf{k}\right| \left|\textbf{B}_{\rm ext}\right|} -\frac{\, \cos \theta \, \nabla_\textbf{x}\, \left|\textbf{B}_{\rm ext}\right|}{ \left|\textbf{B}_{\rm ext}\right|} .
\end{align}

\subsection{Neutron Stars}

It is interesting to plot the conversion probabilities in a realistic astrophysical setup. Neutron stars have been the subject of considerable attention as targets for indirect detection of axions \cite{pshirkov2009,hook2018,ref:NS-Japan,Battye_2020,Battye:2021xvt,Battye:2023oac,Witte:2021arp,Witte:2022cjj,Leroy:2019ghm,foster2020,FosterSETI2022,darling2020apj,darling2020prl}. Here, the plasma profile can be approximated by the Goldreich-Julian model \cite{goldreich1969}. This model consists of an oblique rotating magnetic dipole \cite{Petri:2016tqe}, whose components, in polar coordinates, read
\begin{align}
B_{\rm R} &= B_0 \left(\frac{R}{r}\right)^3 \left(\cos\alpha\cos\theta_{\rm NS} + \sin\alpha\sin\theta_{\rm NS}\cos\psi\right)\,, \nonumber  \\
B_\theta &= \frac{B_0}{2} \left(\frac{R}{r}\right)^3 \left(\cos\alpha\sin\theta_{\rm NS} - \sin\alpha\cos\theta_{\rm NS}\cos\psi\right)\,, \nonumber \\
B_\phi &= \frac{B_0}{2} \left(\frac{R}{r}\right)^3 \sin\alpha \sin\psi \, ,
\end{align}
where $\alpha$ is the angle between the rotation axis and magnetic dipole moment and $\psi(t) = \varphi_{\rm NS} - \Omega \,t$, where $(\theta_{\rm NS}, \varphi_{\rm NS})$ are polar coordinates defined with respect to the rotational axis. The neutron star has a surface magnetic field $B_0$, radius $R$ and rotational frequency $\Omega = 2\pi/P$, where $P$ is the period. The Goldreich-Julian model gives the density of charge carriers as
\begin{align}
n_{\mathrm{GJ}}(\mathbf{r}) = \frac{2\, \boldsymbol{\Omega} \cdot \mathbf{B}}{e} \frac{1}{1 - \Omega^2 \,r^2\, \sin^2 \theta_{\rm NS}}\,,
\label{eq:nGJ}
\end{align}
where $\boldsymbol{\Omega} = \Omega \hat{z}$ is the constant neutron star rotation vector. Neglecting the relativistic terms in the denominator, we arrive at
\begin{align}\label{eq:GJDensity}
n_{\mathrm{GJ}} = \frac{B_0 \Omega}{2 \, e} \left( \frac{R}{r}\right)^3\left[
\cos \alpha + 3 \cos \alpha \cos(2 \theta_{\rm NS}) + 3 \sin \alpha \cos \psi  \sin 2 \theta_{\rm NS}
\right]   .
\end{align}

\begin{figure}
	\centering
	\includegraphics[width=0.8\textwidth]{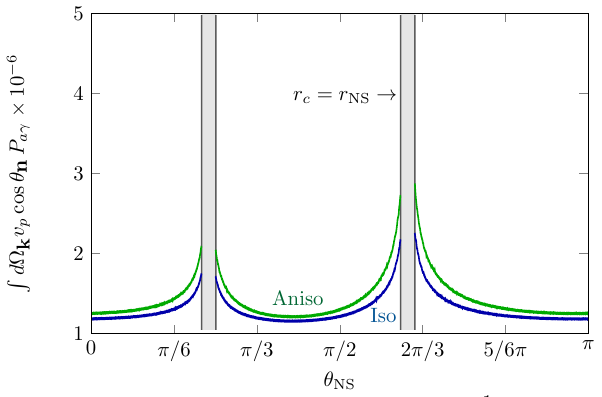}
	\caption{The integrated conversion probability $\braket{P_{a \gamma}} = \int {\rm d}\Omega_\textbf{k} v_p \cos \theta_{\textbf{n}}P_{a \gamma}$ evaluated on the critical surface $r=r_c(\theta_{\rm NS},\varphi_{\rm NS})$ given by $m_a\simeq \omega_{\rm p}$. We chose $\varphi_{\rm NS} = \pi$, $\theta_m=0.4$, $t=0$, $m_a = 10 \mu {\rm eV}$ and $B = 10^{14} {\rm G}, \Omega = 1 {\rm Hz}$, $R= 10{\rm km}$ with $\omega = m_a(1 + v_a^2)^{1/2}$, $v_a= 220 {\rm km/sec}$ and $g_{a \gamma \gamma} =10^{-14}$ GeV. We display in blue the 1D isotropic conversion probability \eqref{eq:IsoProb} and in green, the new result in Eq.~\eqref{eq:PStrongB} of this work. Gray regions show where the production surface would be inside the star, where no photons are produced from axions.}
	\label{fig:Integrated Conversion Probability}
\end{figure}

\begin{figure}[!ht]
	\centering
	\includegraphics[width=0.75\textwidth]{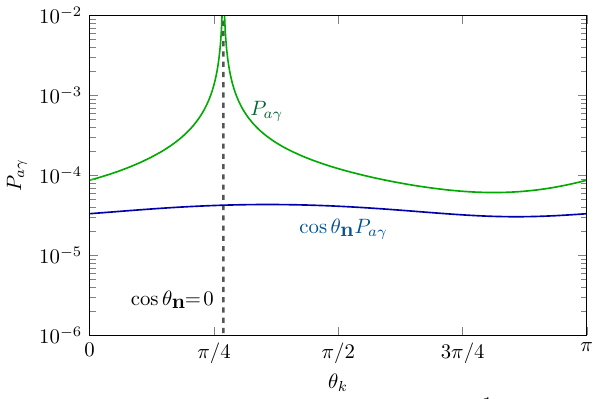}
	\caption{Angular dependence of the anisotropic conversion probability in Eq.~\eqref{eq:PStrongB} (green), as a function  of polar momentum angle $\theta_k$ of the axion. We fixed $\phi_k = 0.2$ and neutron star angles $\theta_{\rm NS} = 0.5$ and $\varphi_{\rm NS} = 0.6 \pi$. Other values are as in Fig.~\ref{fig:Integrated Conversion Probability}. The blue curve shows how the divergences in the probability are regulated by the phase-space measure, as discussed in Sec.~\ref{sec:ContinuitySec}. The gray dashed line indicates the pole in the conversion probability at $\cos \theta_\textbf{n} =0$, when the incoming axion momentum is tangent to the conversion surface. 
 }
	\label{fig:CP_Plots}
\end{figure}

Next, we assume an electrosphere model, in which the magnetosphere is separated into positive and negative charges with mass $m_e$, so that the plasma frequency is given by $\omega_{\rm p} = \sqrt{ e^2 \left| n_{\rm GJ} \right| /m_{\rm e} }$. For a fixed frequency $\omega$, the conversion surface on which $k^\mu_a = k_\gamma^\mu$ is then defined by
 \begin{equation}\label{eq:resCondition}
 	\omega_p^2 = \frac{m_a^2 \omega^2}{m_a^2 + k^2 \sin \theta^2} \simeq m_a^2,
 \end{equation}
 where $\theta$ is the angle between $\textbf{B}$ and $\textbf{k}$, not to be confused with the polar angle $\theta_{\rm NS}$.
The final approximation in Eq.~\eqref{eq:resCondition} holds for non-relativistic axions with $k \ll m_a$. In this limit, all the kinematic surfaces in the foliation $\Sigma_\textbf{k}$ in the right-hand side of Eq.~\eqref{eq:ContinuityFinal} become independent of $\textbf{k}$ and collapse to a single surface $\Sigma$, given by $\omega_{\rm p} = m_a$. It is useful to define a coordinate system for the vector $\hat{\textbf{k}}$ given by
 \begin{equation}
	\hat{\textbf{k}} = \sin \theta_k \cos \phi_k \hat{\textbf{x}} + \sin \theta_k \sin \phi_k \hat{\textbf{y}} + \cos \theta_k \hat{\textbf{z}} \, ,
 \end{equation}
 where $\hat{\textbf{x}}, \hat{\textbf{y}}$ and $\hat{\textbf{z}}$ are defined by the Cartesian axis related to the polar coordinates $(r,\theta_{\rm NS},\varphi_{\rm NS})$ and $(\theta_k,\phi_k)$ give the direction of propagation of the axion.  We can then define an averaged probability over all $\hat{\textbf{k}}$ directions by
 \begin{equation}\label{eq:PAvg}
 	\braket{P_{a \gamma }} \equiv \int {\rm d} \Omega_{\textbf{k}} \, v_p\, \cos \theta_\textbf{n} \, P_{a \gamma}  ,
 \end{equation}
 where ${\rm d}\Omega_\textbf{k}$ is the solid angle appearing in the momentum phase-space measure ${\rm d}^3 \textbf{k} ={\rm d}k k^2 {\rm d} \Omega_\textbf{k} $, and $\theta_\textbf{n}$ is the angle between the phase velocity $\textbf{v}_p = \textbf{k}/\omega$ and the unit normal $\textbf{n}$  to the conversion surface $\Sigma_\textbf{k}$. The normal $\textbf{n}$ is parallel to  $\nabla_\textbf{x} E_\gamma$. The factor $\cos \theta_\textbf{n}$ in the definition \eqref{eq:PAvg} has been pulled out from the phase-space measure in Eq.~\eqref{eq:ContinuityFinal} to regulate $P_{a \gamma }$. The resulting equation \eqref{eq:PAvg} is therefore a function of the remaining phase-space coordinate $\textbf{x}$ (as well as k = |\textbf{k}|, but this is fixed by kinematics once we specify a frequency $\omega$). The radial coordinate $r$ can also be eliminated by using the resonant condition $\omega_p = m_a$, which defines a critical surface $r = r_c(\theta_{\rm NS},\varphi_{\rm NS})$. Hence, on this surface, $\braket{P_{a \gamma}}$ is a function of the polar coordinates $\theta_{\rm NS}$ and $\varphi_{\rm NS}$.  This function is displayed as a function of polar angle $\theta_{\rm NS}$ in Fig.~\ref{fig:Integrated Conversion Probability}. Notice that, for certain angles $(\theta_{\rm NS}, \varphi_{\rm NS})$, the conversion surface penetrates the star, and no production occurs, as shown by the excised gray regions. In Fig.~\ref{fig:CP_Plots}, we also show how divergences in the conversion probability $P_{a \gamma}$ of Eq.~\eqref{eq:PStrongB} are regulated by the flux angle $\cos \theta_\textbf{n}$, which appears naturally in the phase-space measure, as discussed in Sec.~\ref{sec:ContinuitySec}. Note that due to geometry, varying $\theta_k$ between $(0,\pi)$ does not necessarily mean $\cos \theta_\textbf{n}$ fully ranges over $(0,1)$.   

The results presented here should be compared with those of Ref.~\cite{Millar2021}, wherein the Authors analyse the equations of axion electrodynamics directly. The resulting expression for the conversion probability differs from our result~\eqref{eq:PStrongB}, as obtained from the transport equation for the photon distribution functions. Understanding the origins of the discrepancies between these approaches, and any qualitative and quantitative differences in the conversion probabilities, is clearly of great interest, not least from the perspective of analysing how kinetic theory yields results consistent with the classical wave equations. We leave this for future work. We remark, however, that the expression for the conversion probability in Ref.~\cite{Millar2021} requires the introduction of an arbitrary cutoff (see Fig.~3 of Ref.~\cite{Millar2021}) to regulate divergences. By contrast, our expression~\eqref{eq:PStrongB} does not lead to divergent results, and the conversion probability is rendered finite (see Sec.~\ref{sec:ContinuitySec}) by the phase-space measure without further input, see Fig.~\ref{fig:CP_Plots}.

\section{Parity-Even Couplings}\label{sec:parityEven}

Before concluding, we remark on the ease by which this approach can be generalised to couplings of parity-even scalars to the Maxwell term. Such couplings often arise via triangle diagrams in theories with gauge-singlet scalars that couple directly to the Standard Model fermions or mix with the Standard Model Higgs, or in certain scalar-tensor theories of gravity that involve additional gauge-singlet scalar fields that are non-minimally coupled to the Ricci scalar.  In the latter case, the (inverse) Primakoff effects have been used directly to constrain such models~\cite{Burrage:2008ii, Pettinari:2010ay}.

In these cases, the relevant interaction takes the form 
\begin{equation}
    \mathcal{L}\supset -
     \frac{g_{\phi\gamma\gamma}}{4}
    \phi F_{\mu\nu}F^{\mu\nu}= - 
    \frac{ g_{\phi\gamma\gamma}}{2}
   \phi\left(\mathbf{B}^2-\mathbf{E}^2\right).
\end{equation}
In the presence of a background magnetic field $\mathbf{B}_{\rm ext}$, this leads to
\begin{equation}
    \mathcal{L} \supset 
    g_{\phi\gamma\gamma}\mathbf{A}\cdot\left(\mathbf{B}_{\rm ext}\times \nabla\phi\right),
\end{equation}
wherein we have neglected terms that involve gradients of the external magnetic field. This is justified since we have in mind fields $\phi$ whose wavelength is much shorter than that of the background field, thereby allowing us to neglect $\partial_i \textbf{B} \ll \partial_i \phi $. This leads to a contribution to the photon self-energy given by
\begin{equation}
\Pi_{ij}^<(k,X)=
g_{\phi \gamma \gamma}^2
\left(\mathbf{B}_{\rm ext}\times \mathbf{k}\right)_i\left(\mathbf{B}_{\rm ext}\times \mathbf{k}\right)_j2\pi\delta(k^2-m_{\phi}^2)f_{\phi}(k,X),
\end{equation}
in place of the expression in Eq.~\eqref{eq:SelfEnergy} for the parity-odd scalar coupling. This leads to a conversion probability 
\begin{equation}
	P^{\rm CP-even}_{ \phi \gamma  } =   
 \frac{  \pi g_{\phi \gamma \gamma}^2 \big| \hat{\boldsymbol{\varepsilon}}\cdot (\textbf{B}_{\rm ext} \times \textbf{k}) \big|^2	}{  E^2_\gamma\left| \textbf{v}_p \cdot \nabla_\textbf{x} E_\gamma(\textbf{k},\textbf{x})\right|} \cdot \frac{U_E}{U}.
\end{equation}
Applications of the present approach for constraining the couplings of parity-even scalars to the Maxwell term are left for future work. 

\section{Conclusions}\label{sec:conclusions}

In this work, we have provided a robust derivation for the conversion of axions to photons in a strongly magnetised plasma with arbitrary 3D geometry. In fact, our calculation is valid in any medium for which the dispersion relations and polarisations of photons are known, and where the photon mean free path is large compared to the spatial extent of the production region. This calculation has been wanting for many years, since the mechanism of resonant photon production in neutron stars was first re-popularised some half a decade ago \cite{hook2018,ref:NS-Japan}. Indeed, since then, there has been an active programme of both theoretical and observational work carried out to understand and search for these signals \cite{Battye_2020,Battye:2021xvt,Battye:2023oac,Witte:2021arp,Witte:2022cjj,Leroy:2019ghm,foster2020,FosterSETI2022,darling2020apj,darling2020prl}. It is striking that despite this activity, a reliable calculation of the most fundamental part of this story --- the production mechanism of photons themselves --- has remained elusive.

What has been lacking until now is the identification of an appropriate theoretical language to describe axion-photon conversion in such a way that the complexities of 3D geometries, multiple polarisations, background inhomogeneities and differing dispersion relations of axions and photons (amongst other problems) can be dealt with. In the present work, we have shown how kinetic theory --- that is, a theory describing the phase-space of a stochastic ensemble of particles and their collisions --- is a powerful formalism to describe axion-photon mixing. 

To derive a set of transport equations (also called, Boltzmann, kinetic or radiative transfer equations), we deployed powerful techniques from non-equilibrium quantum field theory. This allowed us to derive the transport equations (see Eqs.~\eqref{eq:TransportBornApprox} and \eqref{eq:BoltzmannExplicit}), which describe the propagation and production of photons directly. Without the tools of non-equilibrium quantum field theory at our disposal, the form of these equations would simply have to be intuited on the basis of generalising other formulae for bare photons in vacuum. Furthermore, deriving the correct form of the source term arising from axion production would have been challenging without these techniques. That being said, rather pleasingly, the final answer, having been derived on the basis of a rigorous calculation, then has an immediate physical interpretation in terms of a classical Boltzmann equation for particles with a classical Hamiltonian $\mathcal{H}$. The source term (collisional piece) resembles the standard source term for gauge-boson production \cite{Caron-Huot:2006pee,Gervais:2012wd,Bellac:2011kqa,Kapusta:2006pm}.

These equations give a full 3D description of the system and completely encode the kinematics and dispersion relations for axions and photons, which are in general distinct away from the resonance. At first sight, solving the transport equations in 8-dimensional phase space looks unwieldy. However,  by recasting these equations in terms of characteristics (i.e., the photon rays arising from a system of Hamilton's equations), we were immediately able to probe the 3D structure of the system, capture the refraction of photons and, at the same time, the conversion from axions to photons within a 3D environment. By evolving the system along photon worldines, we were then able to solve for the outgoing phase-space density of Langmuir-O modes and derive a ``conversion probability'', which gives the ratio of photon to axion distribution functions. Rather satisfyingly, the conversion probability reproduces the previously derived limit for axion-photon conversion in an isotropic plasma, and the generalisation to 3D has a similar structure when written in terms of polarisation vectors. Our final answer is free from divergences within the limits of perturbation theory, without the need to impose arbitrary cutoffs.

As discussed in the main text, the expression for the conversion probability~\eqref{eq:PStrongB} obtained in this work from kinetic theory differs from that obtained in an existing analysis of the classical wave equations of axion electrodynamics, see Ref.~\cite{Millar2021}. Moreover, we have seen that the divergences appearing in our conversion probability density are rendered finite by the phase-space measure in the integrated conversion probability; specifically, by the flux angle between the incoming axion momentum and the conversion surface $\Sigma_\textbf{k}$. This occurs without the imposition of a separate cutoff, as was necessary in the analysis of Ref.~\cite{Millar2021}. Notwithstanding this difference, it is clear that a detailed and critical comparison of these two approaches and the resulting expressions for the conversion probability is warranted in order to reach a consensus on the correct description of resonant photon production. This will require further analytical and numerical studies beyond the scope of the present work, and it may be presented elsewhere.

We now discuss some other conclusions of this work, which relate to questions raised elsewhere in the literature.

\begin{itemize}
	\item \textit{``Dephasing'' does not occur}\\
	Previous works \cite{Witte:2021arp} have considered whether the conversion probability will experience suppression due to photon refraction relative to the axion which sources it, owing to the two particles having different dispersion relations away from the resonance. This speculative effect was termed \textit{dephasing} \cite{Witte:2021arp} and was parametrised by an \textit{ad hoc} phenomenological factor to suppress the probability according to the amount of refraction. This was done to make conservative estimates for the conversion probability lest the 1D answer overestimate the true conversion probability.
    
	In this work, we were able, for the first time, to account fully for the refraction of photons when computing the conversion probability. Indeed, since our kinetic equations apply to any 3D environment, and the method of characteristics we used to solve them precisely captures the refracted photon worldlines, the final conversion probability \textit{must} contain (subject to the approximations used) all the relevant information concerning 3D effects.  Equation \eqref{eq:ProbabilityPhysical} should therefore fully account for such effects. How then is photon refraction contained in this formula?
    
	The answer is that it is \textit{already} contained in the factor $1/|\nabla_{\textbf{x}} E_\gamma|$ appearing in the probability (see Eq.~\eqref{eq:PStrongB}). On the one hand, this term can be thought of as capturing the physical width of the resonance, in that it quantifies the size of spatial gradients in the background. In this sense, it defines an effective conversion length $L_c \propto /|\nabla_{\textbf{x}} E_\gamma|^{1/2}$ over which the resonance persists. However, this term also completely captures the refraction of the photon, and occurs directly from the force term $\partial_\textbf{x}\mathcal{H}$ in the Boltzmann equation. Thus, $ 1/|\nabla_{\textbf{x}} E_\gamma|^{1/2}$ can also be thought of as capturing a generalised notion of a radius of curvature $\mathcal{R} \propto 1/|\nabla_{\textbf{x}} E_\gamma|^{1/2}$ of the photon due to refraction, which quantifies the degree to which the photon is accelerated as it passes through the resonance. This is true whether the photon is accelerated parallel or transverse to the sourcing axion, so that it is more helpful to think of the conversion probability as scaling as $P_{a \gamma} \propto \mathcal{R}^2$ with $1/\mathcal{R}$ proportional to the acceleration experienced by the photon.
    
	Hence,  the size of background gradients describes the physical width of the resonance but, at the same time, quantifies the amount of refraction (acceleration) experienced by the photon. The two effects are one and the same, simultaneously encoded in the factor $1/|\nabla_\textbf{x} E_\gamma|$. Wider resonances correspond to more weakly refracted photons and vice versa.  

	Finally, one might question whether some of the approximations that we have made could be the reason why such dephasing is not present in our result. Let us therefore catalogue the main approximations made, of which there are three. They are: $(i)$ the quasiparticle approximation in which decay or absorption of photons is neglected in the left-hand side of the Boltzmann equation; $(ii)$ the truncation of the gradient expansion, i.e., the WKB approximation; and $(iii)$ the Born approximation, in which we work to leading  order in $g_{a \gamma \gamma}$. Clearly $(i)$ has nothing whatsoever to do with refraction. Similarly $(iii)$ is not relevant, since dephasing was claimed to happen to leading order in $g_{a \gamma \gamma}$. Whether or not dephasing occurs at higher order in perturbation remains to be seen.

	Thus, the only tentative avenue for question would be our truncation of the gradient expansion $(ii)$, employed in Eq.~\eqref{eq:DiamondExpansion}. However, truncation at this order reproduces all the required classical equations of kinetic theory. Indeed, this truncation is precisely the WKB approximation, in which particles emerge as a consequence of assuming fluctuations in fields occur at wavelengths much smaller than the characteristic scales of background inhomogeneities. The higher-order terms in the gradient expansion are therefore highly  suppressed by $1/(k L) \ll 1$, where $L$ is the variational scale of the background and $k$ is a characteristic momentum of the particles. Hence, any such dephasing cannot meaningfully be hidden in higher-order gradients, since, if such a truncation is forbidden, the WKB approximation is not valid, and it makes no sense to talk about localised particles in phase space in the first place. The whole concept of relative refraction between individual particles then breaks down. We therefore conclude that, at leading order in $g_{a \gamma \gamma}$, our calculation is robust, and all relevant effects of refraction are captured in our formula and encoded in the term  $1/|\nabla_\textbf{x} E_\gamma|$.

	\item \textit{Resonance width and wave equations versus S-matrix elements and kinetic equations}
	\\
	It is interesting to restate the explicit form of the Boltzmann equation quoted in Eq.~\eqref{eq:BoltzmannExplicit} of the main text
\begin{equation}\label{eq:conclusionEquation}
   \partial_k \mathcal{H} \partial_x f_\gamma	- \partial_x \mathcal{H} \partial_k f_\gamma  =   g_{a \gamma \gamma}^2 \big| k \cdot \tilde{F}_{\rm ext} \cdot \varepsilon \big|^2 2 \pi \delta \left(E_\gamma ( \textbf{k},x)^2 - E_\phi(\textbf{k},x)^2 \right) f_\phi .
\end{equation}
From the delta function on the right-hand side, it naively appears that the resonance happens instantaneously when $E_\gamma = E_\phi$. However, if one were to solve classical wave equations for the mixing between the photon field $A_\mu$ and the axion field $\phi$, as was done in, e.g., Refs.~\cite{hook2018,Battye_2020}, one would observe that the photon field takes a finite amount of integration time to grow across the resonance. The typical scale over which the photon is resonantly driven is set by some length, $L_c \propto 1/(\textbf{k} \cdot \nabla E_\gamma)$. How then, should we reconcile these two pictures? The answer lies in the fact that ultimately we are only interested in the \textit{asymptotic} axion and photon states. That is, we require the asymptotic value of $f_\gamma$ along photon worldlines in the kinetic equations, or the asymptotic values of the electromagnetic fields in the classical wave equations. The conversion probability inferred from the wave equations is then captured by the ratio of the amplitudes of the asymptotic ingoing axion field and outgoing photon field. However, the asymptotics of classical field equations are precisely captured by the S-matrix of tree-level quantum field theory (see Fig.~\ref{fig:SelfEnergy}). Furthermore, the Schwinger-Dyson equations introduced in Sec.~\ref{sec:SchwingerDyson} directly generate the collision integrals, and therefore provide a means to read off these S-matrix elements. The physical width of the resonance is then determined by finding the roots of the delta function in Eq.~\eqref{eq:conclusionEquation}, which gives factors proportional to $1/|\nabla_\textbf{x} E_\gamma|$. Hence, although the kinetic equation does not contain information about the transient evolution of the wave equations through the resonance, it does fully capture the asymptotic behaviour of solutions, and therefore provides a powerful and elegant route to computing the conversion probability that would be derived if a correct solution to the classical equations could be sought.

\newpage
    
	\item  \textit{Divergences}\\
	Our final result is free from divergences, which were thought to arise as a consequence either of a breakdown of stationary phase or as a consequence of a breakdown of WKB \cite{Millar2021}. Indeed, in a previous work \cite{Battye:2021xvt}, some of us argued that the naive divergence observed when $\textbf{k}$ is perpendicular to $\nabla _\textbf{x} E_\gamma$ arose from a breakdown of stationary phase due to so-called degenerate stationary points \cite{poston1996catastrophe}. Such divergences occur because the conversion probability $P_{a \gamma}$ is proportional to $1/|\textbf{k} \cdot \nabla E_\gamma|$. However, we have now shown that such divergences are naturally regulated by the phase-space measure, as explained in the paragraph immediately preceding Sec.~\ref{sec:MagnetisedPlasmas}.  Furthermore, our calculation does not actually involve any stationary phase approximation at all. Hence, these ``perpendicular'' divergences do \textit{not} require arbitrary regulation, but are naturally regulated by the phase-space measure. The exception is when $\left| \nabla_\textbf{x} E_\gamma\right|$ itself becomes small. However, this represents a breakdown of perturbation theory, as we discuss below.

	\item  \textit{Breakdown of perturbation theory and the adiabatic limit}\\
	The one divergence that persists in our result occurs when $\left| \nabla_\textbf{x} E_\gamma\right|$ becomes small compared to terms in the numerator. Here, there is no regulatory behaviour from the phase-space measure. Instead, this divergence represents background gradients becoming large, i.e., the width of the resonance becoming large. As such, the resonant interaction can no-longer be treated as a short-lived perturbation, and the conversion becomes adiabatic, as discussed in Ref.~\cite{Battye_2020}. In this case, perturbation theory breaks schematically when
	\begin{equation}
   	\left| \nabla_\textbf{x} E_\gamma \right| \ll  g_{a \gamma \gamma}^2  (  \boldsymbol{\varepsilon} \cdot \textbf{B}_{\rm ext} )^2.
	\end{equation}
	In this adiabatic limit, one must derive a resummed conversion probability involving an expression to all orders in $g_{a \gamma \gamma}$. In the case of 1D conversion, some of us showed in Ref.~\cite{Battye_2020} that the conversion probability for axion-photon mixing in the adiabatic limit was given by resumming the perturbative result $P_{a\gamma } \rightarrow 1 - \exp(-P_{a \gamma })$. This expression is the well-known Landau-Zener result \cite{Brundobler}. Physically, this effect corresponds to a process in which the resonance persists for so long that, inside the resonant region, axions convert into photons and then convert back into axions again, with this process repeating \textit{ad infinitum} to higher and higher orders in $g_{a \gamma \gamma}$.  As pointed out in Ref.~\cite{Witte:2022cjj}, such effects can be important in strongly magnetised media, such as occur around magnetars. In that instance, one would have to derive a generalised Landau-Zener result for 3D mixing valid to all orders in $g_{a \gamma \gamma}$. This would require finding a non-perturbative solution to our kinetic equations, which generalises the Landau-Zener result to 3D. Undertaking such a calculation lies beyond the scope of the present work, but it is of importance if one is to place limits for larger values of $g_{a \gamma \gamma}$ using strongly magnetised stars.
 
\end{itemize}

In summary, we have provided a robust calculation for resonant mixing of photons with axions in 3D media. This can then be directly inserted into numerical routines or analytic calculations. This may help to settle one of the major open theoretical questions in indirect searches for axions in astrophysical environments, leaving mainly astrophysical uncertainties (e.g., magnetosphere modelling) as the remaining area for improvement in the robustness of future indirect axion searches.
\newpage

\acknowledgments

We thank Alex Millar for comments on the draft and alerting us to a referencing error. J.I.M.~is indebted to Sam Witte for many useful discussions and thanks Richard Battye and the University of Manchester for hospitality during the development of this work. He also acknowledges support from an FSR Fellowship and latterly the Science and Technology
Facilities Council (STFC) [Grant Nos.~ST/T001038/1 and ST/X00077X/1]. P.M. is supported by a United Kingdom Research and Innovation (UKRI) Future Leaders Fellowship [Grant No.\ MR/V021974/2] and the Science and Technology
Facilities Council (STFC) [Grant No.\ ST/X00077X/1]. For the purpose of open access, the authors have applied a Creative Commons Attribution (CC BY) licence to any Author Accepted Manuscript version arising.

\section*{Data Access Statement}

No data were created or analysed in this study. Figs.~\ref{fig:Integrated Conversion Probability} and \ref{fig:CP_Plots} can be reproduced using the code found at this  \href{https://doi.org/10.5281/zenodo.10276950}{link}.

\appendix

\section{Kinetic and Constraint Equations}\label{App:kinetic}

Here, we decompose the Kadanoff-Baym equation \eqref{eq:KB1} for the photon into kinetic and constraint equations~\cite{Zhuang:1995jb,Kainulainen:2001cn,Prokopec:2003pj,Prokopec:2004ic}.  The constraint equation yields the on-shell conditions and polarisation vectors for the photons, and the kinetic equation accounts for their dynamics. The objective is to derive the form of the kinetic equation for the polarisation eigenstates of the photons, i.e., Eq.~\eqref{eq:kinetic}. To arrive at this form, we show here that the solutions for the photon Wightman functions in Wigner space take the approximate form (in the narrow width approximation) given by Eq.~\eqref{eq:SpectralAnsatz} of the main text. Equation \eqref{eq:SpectralAnsatz} accounts for the effects of plasma gradients, which refract photons, and also accounts for photon production due to interactions with axions and the background magnetic field.

We shall use Hermiticity properties of the basic two-point functions that hold when the spacetime indices are both either covariant or contravariant. For two-point functions $A^{\mu\nu}$ and $B^{\mu\nu}$ transforming as rank-two (bi)tensors, we therefore introduce the shorthand notation
\begin{align}
    (A\cdot B)^{\mu\nu}=A^{\mu\rho}\eta_{\rho\sigma}B^{\sigma\nu}\,.
\end{align}
Accordingly, for commutators $[\cdot,\cdot]_-$ and anticommutators $[\cdot,\cdot]_+$, we abbreviate
\begin{align}
    [A,B]_\mp=A\cdot B \mp B\cdot A.
\end{align}

In Sec.~\ref{sec:SchwingerDyson}, we have derived from the manifestly causal equation~\eqref{eq:Wightmann1} the form~\eqref{eq:KB1} of the Kadanoff-Baym equation that is commonly used as a starting point for deriving kinetic theory. To proceed, we introduce the concept of a Wigner transform. We define the relative coordinate $s^\mu$ and central coordinate $X^\mu$ via
\begin{equation}\label{eq:Coords}
s^\mu \equiv x^\mu - y^\mu, \qquad X^\mu \equiv\frac{x^\mu + y^\mu}{2}.
\end{equation}
The Wigner transforms of the two-point functions are defined as Fourier transforms with respect to the relative coordinate $s^\mu$. The Wigner transform of any two-point function $F(x, y)$ in $\left\{D^{\lessgtr}, D^{R(A)},\Pi,...\right\}$ is (suppressing Lorentz indices $\mu, \nu$ etc.~hereafter)
\begin{equation}
F(k, X) \equiv \int {\rm d}^4s\;  e^{ik\cdot s} F(X + s/2, X - s/2)\, .
\end{equation}
The resulting transforms are functions of the phase-space variables $(k,X)$; $X$ can be thought of as the position and $k$ the momentum.
The Wigner transform of convolutions of two-point functions $A$ and $B$ satisfies
\begin{equation}\label{eq:Convolution}
\int {\rm d}^4(u-v) e^{ik\cdot(u-v)} \int {\rm d}^4w\; A(u,w)B(w,v) = e^{-i\Diamond{A(k,X)}{B(k,X)}},
\end{equation}
where $X = (u + v)/2$ is the average coordinate and the diamond operator is defined by:
\begin{equation}\label{eq:DiamondDefinition}
\Diamond \left( \cdot \right) \left( \cdot \right)  = \frac{1}{2} \left(\partial^{(1)}_X\cdot\partial^{(2)}_k - \partial^{(1)}_k \cdot \partial_X^{(2)}\right) \left( \cdot \right) \left( \cdot \right) .
\end{equation}
The superscripts $(1)$ and $(2)$ indicate action on the first and second argument in the round brackets. Thus, if we were to perform a Wigner transform of Eq.~\eqref{eq:Wightmann1} or \eqref{eq:KB1}, we would arrive at an infinite series of derivatives arising from background gradients. Note that, when expanding to first order in gradients, we can write
\begin{equation}\label{eq:DiamondExpansion}
e^{-i\Diamond}{A(k,X)}{B(k,X)} = A(k,X)B(k,X) +
\frac{i}{2}
\left\{
A(k,X), B(k,X) \right\} + \mathcal{O}(\partial_X\cdot \partial_k)^2,
\end{equation}
where we made use of the Poisson bracket defined in Eq.~\eqref{eq:PB}, which we will use frequently in the subsequent approximations. We further note that the Wigner transform of the kinetic operator $\mathcal{D}_0$ in Eq.~\eqref{eq:DmunuSpatial} can be written as
\begin{align}\label{eq:AppWigKinetic}
\overline{\mathcal{D}}_0^{\mu\nu}(k,X)={\cal D}_0^{\mu\nu}(k)+
    \frac i2 \frac{\partial}{\partial k^\alpha}  {\cal D}_0^{\mu\nu}(k)\partial^\alpha_X -\frac18 \frac{\partial}{\partial k^\alpha} \frac{\partial}{\partial k^\beta}{\cal D}_0^{\mu\nu}(k)\partial^\alpha_X \partial^\beta_X,
\end{align}
with
\begin{align}\label{eq:Dkin:momentumspace}
    \mathcal{D}_0^{\mu\nu}(k)=-k^2 \eta^{\mu\nu}+k^\mu k^\nu -\frac1\xi P^{\mu\alpha}P^{\nu\beta}k_\alpha k_\beta,
\end{align}
where the subscript $X$ in Eq.~\eqref{eq:AppWigKinetic} indicates the presence of $X$-derivatives that act on $D^{<,>}(k,X)$.

With these specifications, the Wigner transform of the Kadanoff-Baym equation~\eqref{eq:KB1} reads
\begin{align}\label{eq:appKBWIgner}
   \overline{\cal D}_0\cdot D^{<,>} - e^{-i\Diamond}\Pi^H D^{<,>}
    -e^{-i\Diamond}\Pi^{<,>} D^{H}=-\frac i2 e^{-i\Diamond}\Pi^{>} D^{<} +\frac i2 e^{-i\Diamond}\Pi^{<} D^{>},
\end{align}
where the operator $\Diamond$ is defined in Eq.~\eqref{eq:DiamondDefinition}. Following Refs.~\cite{Zhuang:1995jb,Kainulainen:2001cn,Prokopec:2003pj,Prokopec:2004ic}, to obtain kinetic and constraint equations, we take respectively the anti-Hermitian and Hermitian parts of Eq.~\eqref{eq:appKBWIgner}, which yields
\begin{align}
\label{eq:app:kinetic:constraint}
    \bigg[{\cal D}
    ,D^{<,>}\bigg]_\mp
    +\frac i2 \bigg[\partial_{[k} {\cal D}, \partial_{X]} D^{<,>}\bigg]_\pm 
    =&-\frac i2 \bigg[\Pi^>,D^<\bigg]_\pm+\frac i2 \bigg[\Pi^<,D^>\bigg]_\pm + \bigg[\Pi^{<,>},D_{\rm H}\bigg]_\mp\notag \\
    & 
    + \frac{1}{4}\bigg[\partial_{[k} \Pi^{>},\partial_{X]}D^<\bigg]_\mp  - \frac{1}{4}\bigg[\partial_{[k} \Pi^{<},\partial_{X]}D^>\bigg]_\mp \notag \\
    &+ \frac{i}{2}\bigg[\partial_{[k} \Pi^{<,>},\partial_{X]}D_{\rm H}\bigg]_\pm  \, , 
\end{align}
where square brackets on the indices $k$ and $X$ indicate antisymmetrisation defined by
\begin{equation}
\partial_{[k}A \partial_{X]} B = \partial_{k} A \partial_X B- \partial_{X} A \partial_{k} B\, ,
\end{equation}
for functions $A$ and $B$. We also gathered those terms in Eq.~\eqref{eq:appKBWIgner} involving $\Pi_{\rm H}$ and $\mathcal{D}_0$ into the operator
\begin{equation}
    \mathcal{D}^{\mu \nu}(k,X) =  -k^2g^{\mu\nu} + k^\mu k^\nu - \frac{1}{\xi} P^{\mu\alpha} P^{\nu\beta} k_\alpha k_\beta - \Pi_{_{\rm H}}^{\mu \nu},
\end{equation}
quoted in Eq.~\eqref{eq:DRInverse} of the main text. In Eq.~\eqref{eq:app:kinetic:constraint}, the upper sign is for the kinetic (anti-Hermitian) equation and the lower sign is for the constraint (Hermitian) equation, and we have made use of the fact that in Wigner space, the functions $D^{<,>}$, $\Pi^{<,>}$, $D_{\rm H}$ and $\Pi_{\rm H}$ are Hermitian.  Note that, for consistency at this preliminary stage in the calculation, we have retained all terms up to first order in gradients in Eq.~\eqref{eq:app:kinetic:constraint}. At this point, a few remarks about this equation are in order:
\begin{itemize}
\item 
Recall first that (as explained in the main text), in the absorptive self-energies $\Pi^{<,>}$, we only include the production and absorption of photons from and into\footnote{In the present work, since we work in the Born approximation, re-conversion of photons into axions will not be considered, but we retain these terms for completeness in the early stages of our calculations.} axions induced by the background magnetic field, which is the main effect of interest for the present work. I.e., we will replace $\Pi^{<,>}\to \Pi_{\rm ax}^{<,>}$, where $\Pi_{\rm ax}^{<,>}$ is given in Eq.~\eqref{eq:Piax}.

By contrast, we exclude plasma electrodynamics effects $\Pi_{\rm pl}^{<,>}$ that are not of our primary interest. The leading electromagnetic contribution to $\Pi_{\rm pl}^{<,>}$ would arise from Thomson scattering and is of order $\alpha^2$, with $\alpha$ being the Sommerfeld constant. We assume that the attenuation caused by Thomson scattering can be neglected when predicting the signal from axion-to-photon conversion at leading order. Other absorptive effects, such as the cyclotron resonance~\cite{Witte:2021arp}, occur in regions far from where axion-photon conversion happens.

We remark that when the mean free path of the photons is not large compared to the spatial extent of the plasma, the photon distribution functions will develop a thermal contribution as a result of scatterings with the plasma. While not relevant to our present discussions, such contributions can readily be treated within the framework described in this work.
\item
For the dispersive self-energy $\Pi_{\rm H}$, we drop contributions involving axions.  That is, we do not include corrections to the photon dispersion relation arising from the axion background, since we assume these are sub-dominant compared to QED effects. Corrections to the photon dispersion relation arising from axions are $\mathcal{O}((g_{a\gamma\gamma}\mathbf B_{\rm ext}/E_\gamma)^2)$ (see, e.g., Eq.~(53) of Ref.~\cite{Battye_2020}) and so the neglect of dispersive corrections from axions relative to plasma effects is justified whenever $(g_{a\gamma\gamma}\mathbf B_{\rm ext}/E_\gamma)^2 \ll \alpha$. This is, of course, consistent with the perturbative treatment of $g_{a \gamma \gamma}$ used throughout this paper. This conclusion will change in regions where $g_{a \gamma \gamma} \textbf{B}_{\rm ext}/E_\gamma$ becomes large (see the discussion in Sec.~\ref{sec:conclusions}), leading to a breakdown of perturbation theory that requires the use of fully resummed mass-shell relations and a Landau-Zener-like formula for the conversion probability to all orders in $g_{a\gamma\gamma}$ \cite{Battye_2020}. We do not consider this scenario in the present work. 
\item
Moreover, terms involving gradients of $\Pi_{\rm H}$ contribute to the derivatives of the effective Hamiltonian $\mathcal{H}$ shown in Eq.~\eqref{eq:Liouville}. These correspond to the advection and force terms appearing on the left-hand side of Eq.~\eqref{eq:BoltzmannTime}. These terms equivalently describe the refraction of photons as they pass through the plasma, as embodied by Hamilton's equations \eqref{eq:HamiltonEqs}.  In particular, these terms allow us to describe simultaneously the refraction of photons in combination with the right-hand side of Eq.~\eqref{eq:app:kinetic:constraint}, which corresponds to their production. Retention of gradients of $\Pi_{\rm H}$ therefore allows us to fully capture refraction of photons within the production mechanism, addressing issues of so-called ``dephasing'' raised in Ref.~\cite{Witte:2021arp}.

\end{itemize}

Taking the constraint equations (lower sign in Eq.~\eqref{eq:app:kinetic:constraint}) and neglecting the gradients and the collision terms on the right-hand side, we obtain
\begin{align}
\label{eq:app:constraint}
  \Big[{\cal D},D^{<,>}\Big]_+ - \Big[\Pi^{<,>},D_{\rm H}\Big]_+=0.    
\end{align}
For general $D^{<,>}$, i.e., with possible correlations between the physical polarisation states, the commutators do not vanish. They only do so when $D$ is polarisation-diagonal, as we shall assume eventually.

The form~(\ref{eq:app:constraint}) of the constraint equation can be viewed as an inhomogeous equation for $D^{<,>}$ sourced by $\left[\Pi^{<,>},D_{\rm H}\right]_+$. Formally then, the most general solution consists of the sum of the general solution to the homogeneous and a particular solution satisfying the inhomogeneous equation. The source term involves the Hermitian function $D_{\rm H}$ that can be derived, in turn, from the causal, i.e., retarded and advanced propagators $D_{{\rm R},{\rm A}}$ via Eq.~\eqref{eq:AppHermitianDef}.  The equations for the causal propagators on the closed time path involve the spectral self-energy $\Pi_{\cal A}=\Pi^>-\Pi^<$. Now, the $\Pi^{<,>}$ are dominated by absorptive plasma effects $\Pi_{\rm pl}^{<,>}$ that we do not consider in the present work.

We can nonetheless solve the inhomogeneous equation in the zero-width approximation without fully specifying $\Pi_{\cal A}$. For that purpose, we note that  $\Pi^{\cal \mu\nu}_{\cal A}$ must be consistent with the boundary conditions for the causal propagators, i.e., one may approximate its leading effect through an appropriate pole prescription. Further details are not needed because we know that any special solution to the inhomogeneous equation must be consistent with the spectral function
\begin{align}
   \label{eq:app:SpectralFunction}
	\rho(k,X) = D^>(k,X) - D^<(k,X) = \mp 2 \,{\rm{Im} }D_{\rm R/A}(k,X).
\end{align}
The latter can here suitably be obtained in the zero-width approximation.

Namely, we can solve the equation for the retarded and advanced propagators in Wigner space and in the approximation of zero width (i.e., $\Pi^{\cal \mu\nu}_{\cal A}\to 0$) 
\begin{align}
\label{eq:app:DRA}
   ( {\cal D}\cdot D_{\rm R/A} )_{\mu\nu}=-\eta_{\mu\nu}
\end{align}
to obtain
\begin{equation}\label{Eq:app:DR}
	D_{{\rm R/A} \, \mu \nu}(k,X) = \sum_{c} \frac{ \varepsilon^{c}_\mu \varepsilon^{c\, *}_\nu }{\mathcal{H}_c \pm i \eta \sigma(k_0)},
\end{equation}
where the eigenstates and eigenvalues of $\mathcal{D}$ are defined in Eq.~(\ref{eq:OrthonormalBasis}), $\sigma$ is the signum function, and $\eta$ is positive and infinitesimal. Substituting into 
Eq.~(\ref{eq:app:SpectralFunction}), this leads to
\begin{align}\label{eq:app:spectralDecomp}
 \rho_{\mu \nu} = \sum_{c} \varepsilon^{c}_\mu \varepsilon^{c\, *}_\nu   2 \pi \sigma (k_0)\delta(\mathcal{H}_c).
\end{align}

Now, we turn to the solutions of the homogeneous part of the constraint equation, i.e., the possible contributions to $D^{<,>}$ that satisfy
\begin{align}
    \left[{\cal D},D^{<,>}\right]_+=0.
\end{align}
Any Ansatz that we choose for $D^{<,>}$ to solve the kinetic equation must also self-consistently solve the equation above. The homogeneous solutions to the constraint equations can then be constructed as
\begin{align}
\label{eq:app:Dgrle:constraint}
    D^{<,>}_{\mu \nu}(k,X)\propto
    \sum\limits_c \varepsilon_\mu^c\varepsilon_\nu^{*c} 2\pi \delta({\cal H}_c).
\end{align}

Putting the results for the homogeneous and inhomogeneous pieces together, the general solution to the constraint equations in the zero-width approximation and for vanishing correlations among the energy eigenstates is of the form given in Eq.~(\ref{eq:SpectralAnsatz}).

We turn now to the kinetic equations (Eq.~(\ref{eq:app:kinetic:constraint}), upper sign).
In particular, we shall be interested in projecting these equations onto particular eigenstates by contracting Eq.~(\ref{eq:app:kinetic:constraint}) (upper sign) from the left with $\varepsilon^{a*}$ and from the right with $\varepsilon^a$. To do this, we make use of the following results. Firstly, for any Hermitian objects $A$ and $B$, we have
\begin{equation}
 \varepsilon^{a \, *}  \left[ A , B \right]_\mp \cdot \varepsilon^a =  ( \varepsilon^{a \, *}  \cdot A \cdot  B \cdot \varepsilon^a ) \mp \text{H.c.}, 
\end{equation}
so that after projection onto basis vectors, commutators give anti-Hermitian parts and vice versa. Hence, it follows that projections onto anticommutator terms vanish, i.e.,
\begin{align}
    &\varepsilon^{a*}\cdot \Big[ \mathcal{D}, D^<  \Big]_{-} \cdot  \varepsilon^a  =  2 \,\text{Im}\left( \varepsilon^{a*} \cdot \mathcal{D} \cdot D^< \cdot \varepsilon^a \right)
    =0, \nonumber \\
  & \varepsilon^{a*}\cdot  \big[\Pi^{<,>},D_{\rm H}\big]_- \cdot \varepsilon^a = \,\text{Im}(\varepsilon^{a \, *}\cdot \Pi^< \cdot D_{\rm H} \cdot \varepsilon^a)  = 0.
\end{align}
Meanwhile, for terms involving Poisson brackets, we can write
\begin{align}
   \varepsilon^{a \, *} \cdot \left[ \partial_{[k} A, \partial_{X]} B\right]_{\mp} \cdot \varepsilon^{a } &=  \varepsilon^{a \, *} \cdot  \partial_{[k} A , \partial_{X]} B \cdot \varepsilon^{a }  \mp  \varepsilon^{a \, *} \cdot  \partial_{[X} B , \partial_{k]} A \cdot \varepsilon^{a } \nonumber \\
    &= \varepsilon^{a \, *} \cdot  \left\{ A ,  B \right\} \cdot \varepsilon^{a } \mp  \left( \varepsilon^{a \, *} \cdot  \left\{ A ,  B \right\} \cdot \varepsilon^{a }\right)^*\nonumber \\
    &=\varepsilon^{a \, *} \cdot \left\{ A, B \right\} \cdot \varepsilon^a \mp \text{H.c.}\, .
\end{align}
Putting this together, after projection onto basis vectors, the kinetic equation (upper sign of Eq.~\eqref{eq:app:kinetic:constraint}) reads
\begin{align}\label{eq:KineticFull}
	 \varepsilon^{a*}\cdot
 \left\{ \mathcal{D}, D^<  \right\} 
 \cdot\varepsilon^a  
 + \text{H.c.}  
 &=  \varepsilon^{a*} \cdot  \left[ \left(D^> \Pi^< - \Pi^> D^< \right)     +  
 \left\{ \Pi^<, D_{\rm H}  \right\}  \right ]\cdot\varepsilon^a  + \text{H.c.}  \nonumber \\
 &  +  \frac{i}{2} \left[ \varepsilon^{a*} \cdot  \left(   \left\{ \Pi^<, D^>  \right\} -  \left\{ \Pi^>, D^<  \right\}\right )\cdot\varepsilon^a  -  \text{H.c.}  \right],
\end{align}
where we have used the fact that $\mathcal{D}, D^<, \Pi^<$ and $D_{\rm H}$ are Hermitian in applying the above results. 

Let us now count the powers of $g_{ a \gamma \gamma}$ appearing in Eq.~\eqref{eq:KineticFull}. To leading order in $g_{ a \gamma \gamma}$, the right-hand side of Eq.~\eqref{eq:KineticFull} is independent of the photon distributions $f^c$. To see this, note that (restricting to positive energies) we have
\begin{align}
D_{\mu \nu}^< &=  \sum_{c} 2 \pi\varepsilon^{c}_\mu \varepsilon^{c \, *}_\nu  f^c(k,X) \delta(\mathcal{H}_c), \\ 
D_{\mu \nu}^> &=  \sum_{c} 2 \pi\varepsilon^{c}_\mu \varepsilon^{c \, *}_\nu\left[1+f^c(k,X)\right] \delta(\mathcal{H}_c).
\end{align}
Hence, since $f^c = \mathcal{O}(g^2_{a \gamma \gamma})$, to leading order in $g_{a \gamma \gamma}$, $D^> \Pi^<$ is independent of $f^c$ and $\mathcal{O}(g_{a \gamma \gamma}^2)$. This is the main source for photons due to axions. Meanwhile, $D^< \Pi^>$ is order $\mathcal{O}(g_{a \gamma \gamma}^4)$ and so can be neglected. We also see that 
\begin{equation}\label{eq:DH}
    (D_{\rm H})_{\, \mu \nu} = \text{P.V.}  \sum_c  \frac{\epsilon^c_\mu \epsilon^{c \, *}_\nu}{\mathcal{H}_c},
\end{equation}
where $\text{P.V.}$ denotes the principal value part. Clearly, this is independent of $f^c$, so that the Poisson bracket term $\left\{ \Pi^<, D_{\rm H}  \right\}$ is independent of $f^c$. Thus, we can consider the right-hand side of Eq.~\eqref{eq:KineticFull} as being an $\mathcal{O}(g_{a \gamma \gamma}^2)$ source, which is \textit{independent} of the dynamical variable $D^<$ (or more specifically, $f^c$), which is acted upon by the differential operator on the left-hand side. Putting these arguments together, and using steps outlined in Sec.~\ref{sec:SchwingerDyson} and Appendix \ref{App:Poisson}, we can then write the kinetic equation \eqref{eq:KineticFull} as
\begin{align}\label{eq:KineticDistributionForm}
\delta(\mathcal{H}_c)  \left\{\mathcal{H}_c , f_c \right\}  &= \varepsilon^{c \, *} \cdot \Big( (1+ f_c) \Pi^<  - \Pi^> f_c\Big) \cdot \varepsilon^c  \delta(\mathcal{H}_c)  \nonumber \\
&+\varepsilon^{c*} \cdot    
 \left\{ \Pi^<, D_{\rm H}  \right\}  \cdot\varepsilon^c  + \text{H.c.}  \nonumber \\
 &  +  \frac{i}{2} \left[ \varepsilon^{c*} \cdot   \left( \left\{ \Pi^<, D^>  \right\} -  \left\{ \Pi^>, D^<  \right\}   \right)\cdot\varepsilon^c - \text{H.c.} \right]. 
\end{align}
The general solution to this equation is given by the sum of solutions from sources corresponding to each term on the right-hand side. Let us therefore decompose the solution as
\begin{equation}\label{eq:AppSolutionDecomp}
    f_c = f^0_c+ \bar{f}_c,
\end{equation}
where $f^0_c$ is sourced by the standard collision terms corresponding to the first line on the right-hand side of Eq.~\eqref{eq:KineticDistributionForm}, and $\bar{f}_c$ is sourced by the remaining Poisson bracket terms on the right-hand side. Ultimately, we want to argue that $\bar{f}_c$ does not contribute to asymptotic solutions. To do this, we sum over polarisation indices $c$, which gives
\begin{align}\label{eq:KineticDistributionFormSum}
\sum_c \delta(\mathcal{H}_c)  \left\{\mathcal{H}_c , \bar{f}_c \right\}  =& - \text{Tr}    
 \left\{ \Pi^<, D_{\rm H}  \right\} +  \text{H.c.}  \nonumber \\
 &- \frac{i}{2} \left( \text{Tr}\left[   \left\{ \Pi^<, D^>  \right\}\right] -  \text{Tr} \left[ \left\{ \Pi^>, D^<  \right\} \right] - \text{H.c.}   \right) ,
\end{align}
where, on the right-hand side. we used the completeness relation 
\begin{equation}
    \sum_c \varepsilon^c_\mu \varepsilon^{ c \, *}_\nu = -\eta_{\mu \nu}\, , 
\end{equation}
to contract indices, which leads to the Poisson bracket traces defined by
\begin{equation}\label{eq:TraceDef1}
    \text{Tr} \left \{A ,B \right\} = \partial_{[k} A^{\mu \nu} \partial_{X ]} B_{\mu \nu},
\end{equation}
for Hermitian tensors $A$ and $B$. Next, we notice that Eq.~\eqref{eq:TraceDef1} can be written as the difference of two 4-divergences by writing
\begin{equation}\label{eq:TraceDef2}
    \text{Tr} \left \{A ,B \right\} = \partial_X \cdot ( \partial_k A^{\mu \nu} B_{\mu \nu}) - \partial_k \cdot ( \partial_X A^{\mu \nu}  B_{\mu \nu})  .
\end{equation}
The Poisson bracket trace $\text{Tr}[\left\{ A, B\right\}]$ is real, for any Hermitian operators $A$ and $B$, so that the second line on the right-hand side of Eq.~\eqref{eq:KineticDistributionFormSum} vanishes. The first term on the right-hand side is then real, giving 
\begin{align}\label{eq:KineticDistributionFormSum2}
\sum_c \delta(\mathcal{H}_c)  \left\{\mathcal{H}_c , \bar{f}_c \right\}  =& - 2 \text{Tr}    
 \left\{ \Pi^<, D_{\rm H}  \right\} .
\end{align}
Hence, upon multiplying Eq.~\eqref{eq:KineticDistributionFormSum2} by $k_0$; integrating Eq.~\eqref{eq:KineticDistributionFormSum} over phase space, following similar arguments laid out in Sec.~\ref{sec:ContinuitySec}; and making use of Eq.~\eqref{eq:TraceDef2}, for a stationary system, we arrive at
\begin{equation}
	\sum_c \int {\rm d}^3 \textbf{k} \int {\rm d}\textbf{A} \cdot \textbf{v}_g \, \omega \bar{f}_c  = - 2 \int {\rm d}^3 \textbf{k} \int {\rm d}\textbf{A} \cdot \nabla_\textbf{k} \Pi^{< \, \mu \nu}   D_{{\rm H}\, \mu \nu} .
\end{equation}
This equation relates the asymptotic flux of the contributions $\bar{f}_c$ to a (surface) source term on the right-hand side.  However, the surface term on the right-hand side vanishes if the surface is chosen to be in an asymptotic region outside compactly supported sources from axions. In other words, the volume can be chosen such that $\Pi^{<,>} = 0$ on its bounding surface, A, so  that the right-hand side vanishes. This means that the flux contribution from $\bar{f}_c$ through an asymptotic enclosing surface vanishes, i.e.,
\begin{equation}\label{eq:APP:fluxSum}
    \sum_c \int {\rm d}^3 \textbf{k} \int {\rm d}\textbf{A} \cdot \textbf{v}_g \, \omega \bar{f}_c = 0.
\end{equation}
Further, since each integral and integrand in Eq.~\eqref{eq:APP:fluxSum} is positive, this sum vanishes if and only if $\bar{f}_c$ vanishes everywhere on the surface. We conclude, therefore, that $\bar{f}_c$ vanishes outside the region of axion sources and therefore does not contribute to asymptotic solutions.  

Hence, from an observational perspective, although the Poisson bracket terms on the right-hand side of Eq.~\eqref{eq:KineticDistributionForm} locally give a contribution to the phase-space density of photons, these terms do \textit{not} contribute to the asymptotic solutions. In particular, these solutions are off-shell (as can be seen from Eq.~\eqref{eq:DH}, which contains no mass-shell condition for the photon) and are active only within the axion source. This presumably corresponds to off-shell transient solutions in the classical axion electrodynamics equations, but these do not propagate to infinity away from axion sources. We emphasise that these arguments are a consequence of the Born approximation used throughout this paper, i.e., they only remain true to leading order in $g_{a \gamma \gamma}$. On-shell and off-shell solutions could of course mix at higher order in $g_{a \gamma \gamma}$. Furthermore, the discussion above means that great care should be taken when identifying asymptotic states when comparing classical wave equations to the kinetic equations used here. 

Since we are interested only in those contributions to the asymptotic photon solutions away from the site of axion conversion, we can effectively isolate those parts of Eq.~\eqref{eq:KineticDistributionForm} which determine $f_c^0$, defined in Eq.~\eqref{eq:AppSolutionDecomp}, as the piece that survives into asymptotic regions away from production. This is tantamount to using an effective kinetic equation
\begin{equation}
	\varepsilon^{a \,*} \cdot \left\{ \mathcal{D}, D^<  \right\} \cdot \varepsilon  + \text{H.c.}=  \varepsilon^{a \, *} \cdot \left(D^> \Pi^< - \Pi^> D^< \right) \cdot \varepsilon^a + \text{H.c.} \, .
\end{equation}
This equation therefore captures those on-shell solutions that can propagate away from the axion source, which is the main result appearing in Eq.~\eqref{eq:KineticEq} of Sec.~\ref{sec:SchwingerDyson}.

\section{Eigenmodes in Strongly Magnetised Plasmas and the Spectral Function}\label{App:Eigenmodes}

In this appendix, we analyse the spectral structure of the operator \eqref{eq:DRInverse}, defined by
\begin{equation}\label{eq:HermOP}
\mathcal{D}^{\mu \nu}  \equiv -k^2g^{\mu\nu} + k^\mu k^\nu - \frac{1}{\xi} P^{\mu\alpha} P^{\nu\beta} k_\alpha k_\beta - \Pi_{\rm H}^{\mu \nu},
\end{equation}
which inverts the retarded Green's function $D_{\rm R}$ according to (see Eqs.~\eqref{eq:RetardedEOM} and \eqref{eq:DRInverse} of the main text)
\begin{equation}
 \mathcal{D}^{\mu \nu} (k,X)  \cdot D_{{\rm R }\, \nu\rho}(k,X)   = -\delta^\mu_\rho.
\end{equation}
In Sec.~\ref{sec:SchwingerDyson}, we showed that these relations give rise to the spectral decomposition presented in Eq.~\eqref{Eq:DR} of the main text. Our purpose here is to demonstrate that, for a strongly magnetised plasma, three of the four eigenvalues $\mathcal{H}_c$ have zeros, which correspond to distinct physical states in the plasma. Importantly, we show that two different eigenvalues do not share a common physical zero. Throughout, we use the same coordinates appearing in Eq.~\eqref{eq:FrameChoice}.

The eigenvalues $\mathcal{H}_c$ of $\mathcal{D} $ for an arbitrary $R_\xi$ gauge cannot be written down analytically. However, their product, which gives the determinant of the matrix, has a simple expression
\begin{equation}\label{eq:Determinant}
	\det \mathcal{D}= - \frac{1}{\xi}\frac{|\textbf{k}|^4}{\omega^2} \prod_{i=1}^3(\omega^2 - \omega_i^2),
\end{equation}
where $i$ labels the eigen-energies $\omega_i$ given by
\begin{align}
 \omega^2_{\rm m t} & =  |\textbf{k}|^2, \label{eq:root1}\\  
\nonumber \\
   \omega^2_{\rm LO} &= \frac{1}{2} \Bigg[\,  |\textbf{k}|^2 + \omega_p^2 + \sqrt{|\textbf{k}|^4  + \omega_p^4 + 2\omega_p^2 |\textbf{k}|^2(1 - 2 \cos^2 \theta)} \, \Bigg], \label{eq:root2}\\
  \nonumber \\
  \omega^2_{\rm Alf\text{\'e}n} &=  \frac{1}{2} \Bigg[ \,  |\textbf{k}|^2 + \omega_p^2 - \sqrt{|\textbf{k}|^4  + \omega_p^4 + 2\omega_p^2 |\textbf{k}|^2(1 - 2 \cos^2 \theta)} \, \Bigg]. \label{eq:root3}
\end{align}
Notice that the prefactor $1/\xi$ in Eq.~\eqref{eq:Determinant} prevents the matrix from being singular. This is because removing the gauge-fixing term by taking $\xi \rightarrow \infty$ allows for a trivial zero eigenvalue solution $\mathcal{D}^{\mu \nu}  k_\nu =0$. This follows from $\Pi_{\rm H}^{\mu \nu} k_\nu = 0$ and the projection piece in the first two terms of Eq.~\eqref{eq:HermOP} annihilating $k_\mu$. We see then that setting the determinant to zero gives the physical eigenmodes.  Importantly, these are independent of $\xi$, as required by the fact that the physical shell structure should be gauge independent.

Recall that, in Eq.~\eqref{eq:SpectralAnsatz} of the main text, we used the spectral decomposition of the Wightman functions
\begin{equation}\label{eq:SpectralDecomp}
 D_{\mu \nu}^< =  \sum_{c} 2 \pi\varepsilon^{c}_\mu \varepsilon^{c \, *}_\nu  \left[ \theta(k_0)f_c +\theta(-k_0)\left(1+f_c\right)\right]\delta(\mathcal{H}_c),
\end{equation}
where we interpreted the $f_c$ as corresponding to the phase-space density of the plasma eigenmode labelled by $c$. To justify this, we need to ensure that the zeros of each $\mathcal{H}_a$ correspond to only one physical state, otherwise different terms in this sum in Eq.~\eqref{eq:SpectralDecomp} will mix different states, spoiling the interpretation of the $f_c$ as corresponding to the occupancy of different physical modes. To do this, we must ensure that the eigenmodes are in one-to-one correspondence with the (zeros of) $\mathcal{H}_a$, so that no two $\mathcal{H}_a$ share a common root. Formally, we should not have $\mathcal{H}_a(\omega_i,\textbf{k}) = 0$ and $\mathcal{H}_b(\omega_i,\textbf{k}) = 0$ for $a \neq b$ for any $\omega_i$.

To prove this is the case for the system above, suppose that we have identified a root $k_0 = \omega_i$ such that, for some $\mathcal{H}_a$, we have
\begin{equation}
	\mathcal{H}_a(\omega_i, \textbf{k}) = 0.
\end{equation}
The question then arises as to whether or not there is another $\mathcal{H}_{b \neq a}$ for which this is also a root. Consider the sum of all possible products of eigenvalue triples
\begin{equation}
   \Sigma_3(k_0) =  \sum_{a_1,a_2,a_3} \mathcal{H}_{a_1} (k_0) \mathcal{H}_{a_2}(k_0) \mathcal{H}_{a_3}(k_0).
\end{equation}
We know that any term in the sum containing $\mathcal{H}_a$ vanishes when placed on the $\omega_i-$shell. Since there are four eigenvalues, there is only one term left on shell, which is the product of the other three eigenvalues, i.e.,
\begin{equation}
\Sigma_3(k_0 = \omega_i )  = \prod_{c \neq a} \mathcal{H}_{c}.
\end{equation}
If this term is non-vanishing,  we know that no other eigenvalues have a common root with $\mathcal{H}_a$. However, we also know that products of eigenvalues can be obtained from the moments of the characteristic polynomial of $\mathcal{D}$ defined by
\begin{equation}
\chi (s) = \det \left( s \, {\rm I} - D^{-1}_{R} \right) \equiv \prod_c (s - \mathcal{H}_c ).
\end{equation}
It follows that by differentiating with respect to $s$ and setting $s=0$
\begin{equation}\label{eq:dchids}
 \chi'(0) = - \sum_{a_1,a_2,a_3} \mathcal{H}_{a_1} (k_0) \mathcal{H}_{a_2}(k_0) \mathcal{H}_{a_3}(k_0) .
\end{equation}
Hence, if we can verify that Eq.~\eqref{eq:dchids} is non-vanishing when $k_0 =\omega_i$ for each $\omega_i$, we have shown that no two eigenvalues $\mathcal{H}_c$ share a common root. We find, using the same coordinates as Eq.~\eqref{eq:FrameChoice} and the calculations given in Appendix \ref{Appendix:CovariantGauge}, that the eigenmodes satisfy
\begin{align}\label{eq:CharactersticMoments}
\left. \chi'(0)\right|_{k = \omega} & = -\frac{\omega ^4 \omega_p^2 \sin ^2 \theta }{\xi }, \\
\left. \chi'(0)\right|_{k = k_{\rm LO}, k_{\rm Alfen}} & = - \frac{\omega ^4 \omega _p^2 (\omega ^2 - \omega _p^2) (-\omega ^4 + \omega _p^4 \cos ^2 \theta (-2 + \cos ^2 \theta ) + \omega ^2 \omega _p^2 (1 + \cos ^2 \theta )) \sin ^2 \theta }{\xi (\omega ^2 - \omega _p^2 \cos ^2 \theta )^3}. \label{eq:CharactersticMoments2}
\end{align}
These expressions are clearly non-zero in general, meaning that the $\mathcal{H}_c$ do not share common roots. They only vanish for particular values of $\omega_p$ and $\theta$. Clearly, they vanish if $\omega_p = 0$, since all modes become degenerate in vacuum. Then there is the special case $\theta =0$, which corresponds to a level crossing between the mt and the other two modes, where $k_{\rm LO, \text{Alf\'en}} = \omega$. 

\begin{figure}[t!]
	\centering
	\includegraphics[width=0.8\textwidth]{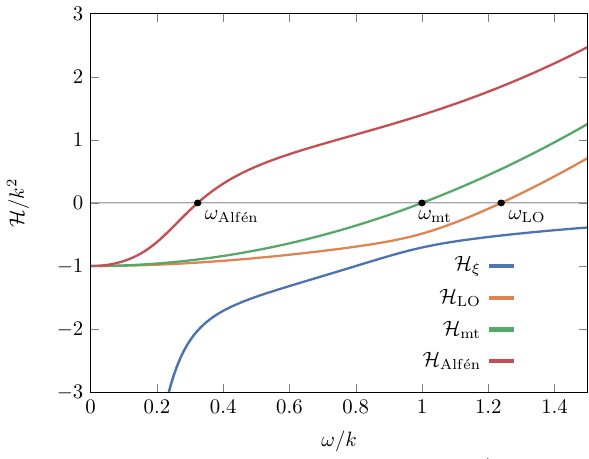}
	\caption{Numerical solutions for the eigenvalues of $D^{-1\, \mu \nu}_{R}$ in Eq.~\eqref{eq:HermOP} as functions of $\omega$. Here, we took $\omega_p/\omega = 0.8 $, $\theta = \pi/3$ and $\xi = 1$. Three of the eigenvalues have roots given by Eqs.~\eqref{eq:root1}, \eqref{eq:root2} and \eqref{eq:root3}. }
	\label{fig:EigenValuePlot}
\end{figure}

Barring level crossings, the eigenvalues $\mathcal{H}_c$ have distinct roots, with each corresponding to one of the mass-shell conditions above. This can be seen in Fig.~\ref{fig:EigenValuePlot}, where we see that the zeros of the $\mathcal{H}_c$ do not overlap and that one of the eigenvalues (which we label $\mathcal{H}_\xi$) is never zero.  The latter arises due to the presence of the gauge term involving $\xi$, which regulates the determinant, preventing it from vanishing off-shell. Hence, there are only three zeros, corresponding to the three physical propagating states in the medium. We have therefore confirmed that the zeros of the $\mathcal{H}_c$ are disjoint, justifying the physical interpretation that each $f_c$ in Eq.~\eqref{eq:SpectralDecomp} corresponds to the distribution functions of \textit{different} physical modes.

\section{Poisson Brackets}\label{App:Poisson}
In this appendix, we derive the Boltzmann equation \eqref{Eq:BoltzmannFull} quoted in the main text. To do so, we show that inserting the spectral Ansatz Eq.~\eqref{eq:SpectralAnsatz} for the photon Wightman function $D^<$ into
\begin{equation}\label{eq:BoltzmannAppendixEQ}
	\varepsilon^{a \,*} \cdot \left\{ \mathcal{D}, D^<  \right\} \cdot \varepsilon^a  + \text{H.c.}=  \varepsilon^{a \, *} \cdot \left(D^> \Pi^< - \Pi^> D^< \right) \cdot \varepsilon^a + \text{H.c.} \, 
\end{equation}
gives rise to Eq.~\eqref{Eq:BoltzmannFull}. Substituting  Eq.~\eqref{eq:SpectralAnsatz} into Eq.~\eqref{eq:BoltzmannAppendixEQ} gives Eq.~\eqref{eq:kinetic} quoted in the main text
\begin{equation}\label{eq:ProjectedBoltzmannAppendix}
 \varepsilon^{a \,*} \cdot  \Big\{ \mathcal{D} ,  \sum_c f_c \delta(\mathcal{H}_c) \varepsilon^c \varepsilon^{c \, *} \Big\} \cdot \varepsilon^a + \text{H.c.} =   \varepsilon^{a \, *} \cdot \Big( (1+ f_a) \Pi^<  - \Pi^> f_a\Big) \cdot \varepsilon^a \, \delta(\mathcal{H}_c) + \text{H.c.} \, ,
\end{equation}
where, as in earlier appendices, we allow ourselves to use $a$ as a polarisation label, since here it cannot be confused with axions. Expanding the Poisson bracket by the product rule and using $\varepsilon^{a \,*} \cdot \varepsilon^c = \delta^{ac}$,  we find
\begin{align}\label{eq:PoissonBraketKinetic}
	&\varepsilon^{a \, *} \cdot  \Big\{ \mathcal{D} ,  \sum_c f_c \delta(\mathcal{H}_c) \varepsilon^c \varepsilon^{c \, *} \Big\}\cdot \varepsilon^{a} \nonumber \\
	&=  (\varepsilon^{a \, *} \cdot \partial_{[k } \mathcal{D} \cdot \varepsilon^{a} \partial_{x]} )\left( f_a \delta(\mathcal{H}_a) \right) \nonumber \\
   &+  \left[ \,f_a \delta(\mathcal{H}_a) (\varepsilon^{a \, *} \cdot \partial_{[k } \mathcal{D} \cdot \partial_{x]} \varepsilon^a)   + \sum_c  f_c\delta(\mathcal{H}_c) (\varepsilon^{a \, *} \cdot \partial_{[k } \mathcal{D} \cdot \varepsilon^c) \, (  \varepsilon^{a } \cdot \partial_{x]} \varepsilon^{c \, *}) \right].
\end{align}
To simplify this expression, we take
\begin{equation}\label{eq:EvalEq}
	\varepsilon^{a \, *}  \mathcal{D}  = \mathcal{H}_a \varepsilon^{a\,*} ,
\end{equation}
which can be differentiated to yield
\begin{equation}\label{eq:EigenEquationDerivative}
	\varepsilon^{a \, *} \cdot \partial_k \mathcal{D} + \partial_k \varepsilon^{a \, *} \cdot \mathcal{D} = \partial_k \mathcal{H}_a  \varepsilon^{a \,*}   +  \mathcal{H}_a \partial_k \varepsilon^{a \, *}  .
\end{equation}
From this, the following three properties are obtained:
\begin{align}
  \textbf{I}:\quad & \varepsilon^{a \, *}  \cdot \partial_k  \mathcal{D}\cdot \varepsilon^a  =  \partial_k \mathcal{H}_a,
 	\\
 	\nonumber \\
  \textbf{II}:\quad  &\delta (\mathcal{H}_c) ( \varepsilon^{a\, *}  \cdot \partial_{[ k}  \mathcal{D} \cdot \varepsilon^c)\,  ( \varepsilon^a \cdot \partial_{x]} \varepsilon^{c \, *})   =0,\label{eq:app:polderII}
 	\\
	\nonumber \\
 \textbf{III}:\quad  &  \varepsilon^{a\,*}  \cdot \partial_{[k} \mathcal{D} \cdot \partial_{x]} \varepsilon^a   = 0 .\label{eq:app:polderIII}
\end{align}
 We prove identities $\textbf{I}$-$\textbf{III}$ at the end of this appendix but, for now, we take them as read and proceed with the calculation. Using these identities, we see that any terms involving derivatives of the polarisation vectors vanish in Eq.~\eqref{eq:PoissonBraketKinetic}, so that
\begin{align}\label{eq:HamiltonianPoisson}
\varepsilon^{a \, *} \cdot  \Big\{ \mathcal{D} ,  \sum_c f_c \delta(\mathcal{H}_c) \varepsilon^c \varepsilon^{c \, *} \Big\}\cdot \varepsilon^{a}
=  \Big\{\mathcal{H}_a , f_a \delta(\mathcal{H}_a) \Big\} =
 \delta(\mathcal{H}_a) \Big\{\mathcal{H}_a , f_a \Big\} \, ,  
\end{align}
where in the last equality we have used the product rule for the Poisson bracket and the trivial fact that $\left\{\mathcal{H}_a , \delta(\mathcal{H}_a) \right\} =0$. Inserting the result \eqref{eq:HamiltonianPoisson} into Eq.~\eqref{eq:ProjectedBoltzmannAppendix} gives
\begin{equation}
\delta(\mathcal{H}_c)  \left\{\mathcal{H}_c , f_c \right\}  = \varepsilon^{c \, *} \cdot \Big( (1+ f_c) \Pi^<  - \Pi^> f_c\Big) \cdot \varepsilon^c  \delta(\mathcal{H}_c)  \, .
\end{equation}
This is precisely the expression \eqref{Eq:BoltzmannFull} of the main text.

As promised, in the remainder of this appendix, we prove the identities $\textbf{I}$-$\textbf{III}$.

\subsection*{Identity I}
To prove this identity, we begin by multiplying Eq.~\eqref{eq:EigenEquationDerivative} from the right by $\varepsilon^a$ and use \eqref{eq:EvalEq} to obtain
\begin{equation}
	\varepsilon^{a \, *} \cdot  \partial_k \mathcal{D} \cdot \varepsilon^a + \mathcal{H}_a  \, (\varepsilon^a \cdot \partial_k \varepsilon^{a \, *})   = \partial_k \mathcal{H}_a 	+  \mathcal{H}_a  \, (\varepsilon^a  \cdot \partial_k \varepsilon^{  a \, * })
\end{equation}
The second terms on each side cancel, leaving one with
\begin{equation}
	\varepsilon^{a \, *} \cdot  \partial_k  \mathcal{D} \cdot \varepsilon^a  =  \partial_k \mathcal{H}_a,
\end{equation}
as required.

\subsection*{Identity II}
We first multiply Eq.~\eqref{eq:EigenEquationDerivative} by a different eigenvector $\varepsilon^c$ from the right, which gives
\begin{equation}
	\varepsilon^{a \, *}  \cdot \partial_k \mathcal{D} \cdot \varepsilon^c + \, \mathcal{H}_c (\varepsilon^c \cdot \partial_k \varepsilon^{a \, *} )  = \delta^{a c} \partial_k \mathcal{H}_a 	+  \mathcal{H}_a  ( \varepsilon^c   \cdot \partial_k \varepsilon^{a \, *}) \, .
\end{equation}
Here, we used the orthonormality condition $\varepsilon^{a \, *} \cdot \varepsilon^c= \delta^{ac}$ in deriving the first term on the right-hand side, and also the identity $\mathcal{D} \cdot \varepsilon^c = \mathcal{H}_c \varepsilon^c$ to obtain the second term on the left-hand side. We can then multiply this by $\delta(\mathcal{H}_c)$, which eliminates the second term on the left-hand side, yielding
\begin{equation}
   \delta(\mathcal{H}_c)  ( \varepsilon^{a \, *}  \cdot \partial_k \mathcal{D} \cdot \varepsilon^c)   =  \delta^{a c}  \delta(\mathcal{H}_c)  \partial_k \mathcal{H}_a 	+  \mathcal{H}_a  \delta(\mathcal{H}_c) \, (\varepsilon^c \cdot  \partial_k \varepsilon^{ a \,*}  ) .
\end{equation}
Finally, we can multiply this by $\varepsilon^a \cdot \partial_x \varepsilon^{c \, *}$ to arrive at
\begin{equation}\label{eq:IdII}
\delta (\mathcal{H}_c) (\varepsilon^{a\, *}  \cdot \partial_{ k}  \mathcal{D} \cdot \varepsilon^c )(\varepsilon^a \cdot \partial_{x} \varepsilon^{c\, *} )   = \delta^{ac} \partial_{k} \mathcal{H}_a \delta (\mathcal{H}_a) (\varepsilon^a \cdot \partial_{x} \varepsilon^{a \, *})  +  \mathcal{H}_a \delta(\mathcal{H}_c) (\varepsilon^{c \, *} \cdot \partial_{k} \varepsilon^a )  (\varepsilon^a \cdot \partial_{x} \varepsilon^{c \, *} ),
\end{equation}
where, in the first term on the right-hand side, we used $\delta^{a c}$ to set $a = c$ in $\varepsilon^a\cdot \partial_x \varepsilon^{c\,*}$. The first term on the right-hand side of Eq.~\eqref{eq:IdII} vanishes, because $\varepsilon^a\cdot \partial_x \varepsilon^{a\, *}  = \partial_x(\varepsilon^a \cdot \varepsilon^{a \, *})/2 =0$, since $\varepsilon^a\cdot \varepsilon^{a\,*} = 1$. This leaves
\begin{equation}
\delta(\mathcal{H}_c) (\varepsilon^{a\, *} \cdot \partial_k \mathcal{D} \cdot \varepsilon^c) ( \varepsilon^a \cdot \partial_x \varepsilon^{c\, *} )  = \mathcal{H}_a \delta(\mathcal{H}_c) (\varepsilon^c \cdot \partial_k \varepsilon^{a \, *}) (\varepsilon^a \cdot \partial_x \varepsilon^{c \, *}).
\end{equation}
However, the second term on the right-hand side is clearly symmetric under interchange of $k$ and $x$. As such, it vanishes by virtue of antisymmetrisation, so that
\begin{equation}
\delta (\mathcal{H}_c) ( \varepsilon^{a\, *}  \cdot \partial_{[ k}  \mathcal{D} \cdot \varepsilon^c )( \varepsilon^a \cdot \partial_{x]} \varepsilon^{c\, *} )   =  0.
\end{equation}
This completes the proof of identity \textbf{II}.

\subsection*{Identity III}
We begin by contracting Eq.~\eqref{eq:EigenEquationDerivative}  with $\partial_x \varepsilon^a$, which gives
\begin{equation}
	(\varepsilon^{a\, *} \cdot \partial_k \mathcal{D} \cdot \partial_x \varepsilon^a) + (\partial_k \varepsilon^{a\,*} \cdot \mathcal{D} \cdot  \partial_x \varepsilon^a)= \partial_k \mathcal{H}_a  (\partial_x \varepsilon^{a}\cdot  \varepsilon^{a\, *} )   +  \mathcal{H}_a (\partial_k \varepsilon^{a\, *}   \cdot \partial_x \varepsilon^a).
\end{equation}
Clearly, the first term on the right-hand side vanishes due again to $\varepsilon^{a \, *}\cdot \partial_x \varepsilon^a  = \partial_x(\varepsilon^{a\, *} \cdot \varepsilon^a)/ 2=0$, from which it follows that
\begin{equation}
  (\varepsilon^{a\, *} \cdot  \partial_k \mathcal{D} \cdot \partial_x \varepsilon^a)   = \mathcal{H}_a (\partial_k \varepsilon^{\, a *}   \cdot \partial_x \varepsilon^a) - (\partial_k \varepsilon^{a\, *} \cdot \mathcal{D}\cdot  \partial_x \varepsilon^a).
\end{equation}
Similarly to the proof of the previous identity, the two terms on the right-hand side are symmetric under interchange of $k$ and $x$, which follows in the second term from the Hermitian nature of $\mathcal{D}$. Hence, upon antisymmetrisation, they both vanish, and we have
\begin{equation}
  \varepsilon^{a \, *}  \cdot \partial_{[k} \mathcal{D} \partial_{x]} \cdot \varepsilon^a   =0 \, ,
\end{equation}
completing the last of the three derivations for the identities above.

\section{Gauge Invariance of the Stored Energy Term}\label{App:GaugeInvce}
In this appendix, we show that $\mathcal{E}^2 \partial_{k_0} \mathcal{H}$ is gauge invariant. This result was required to derive the result \eqref{eq:BoltzmannEMFields}, which recasts the Boltzmann equation in terms of physical electromagnetic fields. To establish the gauge invariance of $\mathcal{E}^2 \partial_{k_0} \mathcal{H}$, we begin with the identity \textbf{I} proved in Appendix \ref{App:Poisson}, namely
\begin{equation}\label{eq:EigenvalueDerivativeAppendix}
\mathcal{E}^* \cdot (\partial_k \mathcal{D}) \cdot \mathcal{E} =  \partial_k \mathcal{H} \mathcal{E}^2,
\end{equation}
where we recall, from the discussion preceding Eq.~\eqref{eq:RatioCollision}, that $\mathcal{E}$ has arbitrary normalisation. Now, consider a gauge transformation \eqref{eq:PolarisationTransformation} to a new polarisation vector
\begin{equation}
	\mathcal{E}'_\mu = \mathcal{E}_\mu - i k_\mu \tilde{\Lambda}.
\end{equation}
The left-hand side of Eq.~\eqref{eq:EigenvalueDerivativeAppendix} then reads
\begin{equation}\label{eq:GaugeTrans}
\mathcal{E}^* \cdot (\partial_k \mathcal{D}) \cdot \mathcal{E} = \mathcal{E}'^{\,*} \cdot (\partial_k \mathcal{D}) \cdot \mathcal{E} '  + \left( i\tilde{ \Lambda}  \left[ \mathcal{E}'^{\, *}  \cdot (\partial_k \mathcal{D}) \cdot k \right] + \text{H.c.}\right) + \big| \tilde{\Lambda}\big|^2 \left[ k\cdot (\partial_k \mathcal{D}) \cdot k\right].
\end{equation}
We can rewrite the second term on the right-hand using 
\begin{equation}\label{eq:EPrimeTerm1}
	\mathcal{E}'^{\, *} \cdot (\partial_k \mathcal{D}) \cdot k =  \mathcal{E}'^{\, *} \cdot \partial_k  (\mathcal{D} \cdot k) - \mathcal{E}'^{\, *} \cdot \mathcal{D} \cdot \eta,
\end{equation}
where $\eta$ is the Minkowski metric, carrying the spare index associated to the derivative. The second term on the right-hand side of the equation above vanishes on-shell, since  $\mathcal{E}^* \cdot \mathcal{D} =0$. Next, since we wish to allow for arbitrary gauges, we should remove the covariant gauge term by taking $\xi \rightarrow \infty$ in Eq.~\eqref{eq:DRInverse}. We then have that, even for off-shell $k_\mu$,
\begin{equation}\label{eq:ArbGauge}
	\mathcal{D}(\xi \rightarrow \infty) \cdot k = 0.
\end{equation}
From this, it follows that the first term on the right-hand side of Eq.~\eqref{eq:EPrimeTerm1} also vanishes, leaving us with
\begin{equation}\label{eq:Zero1}
	\mathcal{E}'^{\, *} \cdot (\partial_k \mathcal{D}) \cdot k  = 0,
\end{equation}
establishing that the second term on the right-hand side of Eq.~\eqref{eq:GaugeTrans} vanishes. Returning now to the last term on the right-hand side of Eq.~\eqref{eq:GaugeTrans}, we see that it can be written as
\begin{equation}\label{eq:Zero2}
	k \cdot (\partial_k \mathcal{D}) \cdot k = k \cdot \partial_k ( \mathcal{D}\cdot k) - k \cdot \mathcal{D}\cdot \eta = 0,
\end{equation}
where both terms vanish by making use again of the relation \eqref{eq:ArbGauge}.  Thus, having established Eqs.~\eqref{eq:Zero1} and \eqref{eq:Zero2}, we see that both the second and third terms on the right-hand side of Eq.~\eqref{eq:GaugeTrans} vanish, leading to
\begin{equation}
\mathcal{E}^* \cdot (\partial_k \mathcal{D}) \cdot \mathcal{E} = \mathcal{E} '^{\, *} \cdot (\partial_k \mathcal{D}) \cdot \mathcal{E} ' .
\end{equation}
This establishes the gauge invariance of $\mathcal{E}^* \cdot (\partial_k \mathcal{D}) \cdot \mathcal{E}$. Furthermore,  since, from Eq.~\eqref{eq:EigenvalueDerivativeAppendix}, this is equal to $\partial_k \mathcal{H} \mathcal{E}^2$, it follows that $\partial_{k_0} \mathcal{H} \mathcal{E}^2$ is also gauge invariant, proving the result quoted in Sec.~\ref{sec:GaugeInvcePhysFields}.

\section{Axion Collision Term}\label{sec:collision}
The axion contribution to the photon Wightman self-energy $\Pi^<$, appearing in Fig.~\ref{fig:SelfEnergy}, can be read off from the interaction vertex \eqref{eq:Lint}. It is
\begin{equation}
	\Pi^<_{\rm ax \, \mu \nu}(x, y) = g_{a \gamma \gamma}^2 \tilde{F}_{\mu \rho}^{\rm ext}(x)  \partial_x^\rho G^<(x ,y) \overleftarrow{\partial}_{y}^\sigma  \tilde{F}^{\rm ext}_{\sigma \nu }(y).
\end{equation}
From Eq.~\eqref{eq:Coords}, we have
\begin{equation}
	\partial_x = \partial_s + \frac{1}{2}\partial_X, \qquad \partial_{y} = - \partial_s + \frac{1}{2}\partial_X,
\end{equation}
so that the Wigner transform sends
\begin{equation}
	\partial_x^\rho G^<(x ,x') \overleftarrow{\partial}_{x'}^\sigma \rightarrow (i k + \partial_X)^\rho ( - i k + \partial_X)^\sigma G^<(k, X).
\end{equation}
The derivatives $\partial_X$ are by assumption sub-dominant in the gradient expansion and can be neglected. Similarly, by Taylor expanding about $s_\mu=0$, we have
\begin{align}
    &\tilde{F}_{\rm ext}^{\mu \nu}(x) = \tilde{F}_{\rm ext}^{\mu \nu}\left(X + \frac{s}{2}\right)= \tilde{F}_{\rm ext}^{\mu \nu}(X)  + s \cdot \partial_X F_{\rm ext}^{\mu \nu}(X) + \cdots , 
\end{align}
where the second term is again gradient suppressed, since, in Wigner space, they are $\mathcal{O}(\partial_k \cdot \partial_X) = \mathcal{O}(1/(k L)) \ll 1 $, where $1/L$ sets the size of background gradients and $k$ is a typical momentum scale. Similarly, we have $\tilde{F}_{\rm ext}^{\mu \nu}(y) \simeq \tilde{F}^{\rm ext}_{\mu \nu}(X)$. Next, using the Wightman function for a spin-zero field \cite{Blaizot:2001nr}
\begin{equation}
	G^<(k,X) = 2 \pi \delta(k^2 - m_\phi^2) \left\{f_\phi(k,X) \theta(k_0) +  \theta(-k_0) \left[1 + f_\phi (-k,X) \right] \right\},
\end{equation}
where $f_\phi$ is the axion distribution function, we arrive at the following expression for the photon Wightman self-energy in Wigner space, quoted in Eq.~\eqref{eq:SelfEnergy}:
\begin{align}
	\Pi^{ < \, \mu \nu}_{\rm ax}(k,X) &= g_{a \gamma \gamma}^2 k_\rho k_\sigma \tilde{F}_{\rm ext}^{\mu \rho}(X) \tilde{F}_{\rm ext}^{\nu \sigma}(X)  2 \pi \delta(k^2 - m_\phi^2) \nonumber\\&\phantom{=}\times\left\{f_\phi(k,X) \theta(k_0) +  \theta(-k_0) \left[1 + f_\phi (-k,X) \right] \right\}.
\end{align}

\section{Conversion Probability in Covariant Gauge}\label{Appendix:CovariantGauge}

In this appendix, we reproduce the same expression for the conversion probability \eqref{eq:PStrongB}, using covariant gauge. Our starting point is the covariant form of the conversion probability in Eq.~\eqref{eq:ProbabilityCovariant}
\begin{equation}\label{eq:AppendixProb}
	P_{a \gamma \gamma } =   \frac{  \pi g_{a \gamma \gamma}^2 \big| k \cdot \tilde{F}_{\rm ext} \cdot \varepsilon \big|^2	}{ E_\gamma \partial_{k_0} \mathcal{H} \left| \textbf{v}_p \cdot \nabla_\textbf{x} E_\gamma(\textbf{k},\textbf{x})\right|} .
\end{equation}
To compute the polarisation 4-vector $\varepsilon$, we require the general polarisation 4-tensor appearing in Eq.~\eqref{eq:DRInverse}.
This can be inferred from the components $\Pi^{ij}_{\rm H}$, which can be read off from the classical theory via Eqs.~\eqref{eq:3polarisationTensor} and \eqref{eq:Permittivity}. The remaining components $\Pi^{0i}_{\rm H}$ and $\Pi^{00}_{\rm H}$ can then be read off \cite{MelroseBookI_QPD2008} by using the transverse condition
\begin{equation}
\Pi_{\rm H}^{\mu \nu} k_{\nu} = 0 ,
\end{equation}
which follows from conservation of the current. From this set of four equations, we can read off $\Pi^{0 i}_{\rm H}$ and $\Pi^{0 0}_{\rm H}$. Working in a frame in which
\begin{equation}\label{eq:FrameChoice}
k_\mu = (\omega , 0 , 0 , k), \qquad  \textbf{B} = (0, -\sin \theta, \cos \theta ), \qquad u^\mu = (1,0,0,0),
\end{equation}
we then infer
\begin{equation}
\Pi^{\mu \nu}_{\rm H} =\left(
\begin{array}{cccc}
  \frac{k^2 \, \omega _p^2 \cos ^2\theta }{\omega ^2} & 0 & \frac{k \,\omega
   _p^2 \sin \theta \cos \theta }{\omega } & -\frac{k \,\omega _p^2 \cos ^2\theta }{\omega } \\
 0 & 0 & 0 & 0 \\
 \frac{k \omega _p^2 \sin \theta \cos \theta }{\omega } & 0 & \omega _p^2\sin ^2\theta  & -  \omega _p^2 \sin
   \theta \cos \theta \\
 -\frac{k \omega _p^2 \cos ^2\theta }{\omega } & 0 & -\omega _p^2\sin \theta \cos \theta  &
   \omega _p^2\cos ^2\theta  \\
\end{array}
\right).
\end{equation}
We can now use this to solve the following eigenvalue equation in a covariant gauge:
\begin{equation}\label{eq:RetardedEOMAppendix}
 \mathcal{D} \cdot \varepsilon = \left[ -k^2g^{\mu\nu} + k^\mu k^\nu - \frac{1}{\xi} P^{\mu\alpha} P^{\nu\beta} k_\alpha k_\beta - \Pi_{\rm H}^{\mu\nu}(k,x) \right] \varepsilon_{\nu } = 0.
\end{equation}
Solving this equation on-shell, we find the polarisation 4-vector of the LO mode to be (up to an overall normalisation)
\begin{equation}\label{eq:PolarisationLO4Vec}
\varepsilon^{\rm LO}_\mu = \left( \sin \theta  \cos \theta ,0,\frac{\left(\omega ^2-\omega _p^2\right) }{\omega _p^2} \sqrt{\frac{\omega
   ^2-\cos ^2 \theta  \omega _p^2}{\omega ^2-\omega _p^2}},0 \right).
\end{equation}
It is straightforward to verify, using $E_i  = F_{0 i } = \partial_0 A_i - \partial_i A_0$, with $A_\mu \propto \varepsilon_\mu^{\rm LO}$, that this reproduces the expression for the physical electric field polarisation \eqref{eq:LO}. Using Eq.~\eqref{eq:PolarisationLO4Vec}, we compute the matrix-element-like term in the numerator of Eq.~\eqref{eq:AppendixProb}, viz.,
\begin{equation}\label{eq:MatrixElt}
  ( k_\mu \tilde{F}^{\mu \nu} \hat{\varepsilon}^{\rm LO}_\nu)^2 = \frac{\omega ^6 \sin ^2 \theta  \left(\omega ^2-\omega _p^2\right)}{\left(\omega ^2- \omega _p^2 \cos ^2\theta  \right) \left[\omega ^4-\omega ^2 \left(\cos ^2\theta
   +1\right) \omega _p^2+\cos ^4\theta  \omega _p^4\right]},
\end{equation}
where $\hat{\varepsilon}^{\rm LO}_\nu$ is the unit 4-vector defined by $\hat{\varepsilon}^{\rm LO}_\nu = \varepsilon^{\rm LO}_\nu/(\varepsilon^{\rm LO\, *}_\rho \eta^{\rho \sigma} \varepsilon^{\rm LO}_\sigma)^{1/2}$. Next, we require the expression for $\partial_{k_0} \mathcal{H}$, which can be written on-shell as
\begin{equation}\label{eq:dHdk}
	\partial_{k_0} \mathcal{H} = \hat{\varepsilon}^{\rm LO \, *} \cdot \partial_{k_0} (\mathcal{D})\cdot \hat{\varepsilon}^{\rm LO}.
\end{equation}
 Upon computing the right-hand side of \eqref{eq:dHdk}, we find
\begin{equation}\label{eq:dHdkExpression}
   \partial_{k_0} \mathcal{H} = \frac{2 \omega  \left(\omega ^2-\omega _p^2\right) \left[\cos ^2\theta \omega _p^2 (\omega _p^2 -2 \omega ^2 ) +\omega ^4\right]}{\left(\omega
   ^2- \omega _p^2 \cos ^2\theta \right) \left[\omega ^4 -\omega ^2 \left(\cos ^2\theta+1\right) \omega _p^2+\cos ^4\theta \omega _p^4\right]}.
\end{equation}
Note that Eqs.~\eqref{eq:MatrixElt} and $\eqref{eq:dHdkExpression}$ are independent of the gauge parameter $\xi$. We now divide Eq.~\eqref{eq:MatrixElt} by Eq.~\eqref{eq:dHdkExpression} and insert the result into Eq.~\eqref{eq:AppendixProb} to obtain
\begin{equation}
   P_{a \gamma} = \frac{\pi}{2} \cdot \frac{ g_{a \gamma \gamma}^2 \left| \textbf{B}_{\rm ext } \right|^2\omega ^4 \sin ^2 \theta }{\cos ^2\theta \omega _p^2 (\omega _p^2 -2 \omega ^2 ) +\omega ^4} \cdot \frac{ 1 }{\left| \textbf{v}_p \cdot \nabla_\textbf{x} E_\gamma(\textbf{k},\textbf{x})\right|} .
\end{equation}
This reproduces exactly Eq.~\eqref{eq:PStrongB}. Thus, we have shown, via a different calculational route and in a Lorentz covariant gauge, the same form of the probability in Eq.~\eqref{eq:PStrongB}, which was derived using temporal gauge and physical electromagnetic fields. Note that the approach in this appendix also circumvents any physical discussion about the stored energy in the modes, which came from substituting $\partial_{k_0} \mathcal{H}$ for the expression \eqref{eq:RM} involving the stored energy in the plasma.  Instead, we have worked directly with the eigenvalues of the matrix in Eq.~\eqref{eq:RetardedEOMAppendix}.

\bibliography{bibliography.bib}{}
\bibliographystyle{JHEP}

\end{document}